%% file: ms-emulateapj.tex
\documentclass{emulateapj}

\usepackage{natbib}
\usepackage{epsfig, color, subfigure}
\usepackage{amssymb,amsmath, epstopdf}
\usepackage{multirow}

\shorttitle{Lensed galaxies}
\shortauthors{Newton et~al.}

\def\apj{ApJ}
\def\nat{Nature}
\def\mnras{MNRAS}
\def\apjl{ApJL}
\def\apjs{ApJS}

\def\paperV{{Paper~V}{}}
\def\paperIX{{Paper~IX}{}}

\def\eg{{\it e.g.}{}}

\def\hst{{\it HST}{}}
\def\sersic{{S\'{e}rsic}{}}

\def\klens{{\sc klens}{}}
\def\galfit {{\sc galfit}{}}
\def\mpfit{{\sc mpfit}{}}
\def\sextractor{{\sc SExtractor}{}}

\def\hstV{{$V_{\rm HST}$}{}}
\def\hstI{{$I_{814}$}{}}
\def\hstH{{$H_{160}$}{}}

\def\zd{z_{\rm d}}
\def\zs{z_{\rm s}}
\def\Reff{r_{\rm eff}}
\def\xd{x_{\rm d}}
\def\yd{y_{\rm d}}
\def\sigmaSIE{\sigma_{\rm SIE}}
\def\thetad{\theta_{\rm d}}

\def\thetas{\theta_{\rm s}}

\def\msun{M_{\odot}}
\def\lsun{L_{\odot}}

\def\secref#1{Section~\ref{#1}}

\begin{document}

\title{The Sloan Lens ACS Survey. XI. 
Beyond Hubble resolution: size, luminosity and stellar mass of compact lensed 
galaxies at intermediate redshift}

\author{Elisabeth~R.~Newton\altaffilmark{1,*}}
\author{Philip~J.~Marshall\altaffilmark{1,2}}
\author{Tommaso~Treu\altaffilmark{1,3}}
\author{Matthew~W.~Auger\altaffilmark{1}}
\author{Raph\"ael~Gavazzi\altaffilmark{4}}
\author{Adam~S.~Bolton\altaffilmark{5}}
\author{L\'eon~V.~E. Koopmans\altaffilmark{6}}
\author{Leonidas~A.Moustakas\altaffilmark{7}}

\altaffiltext{1}{Department of Physics, University of California, Santa Barbara, CA 93106, USA}
\altaffiltext{*}{current address: \texttt{enewton@cfa.harvard.edu}}
\altaffiltext{3}{David and Lucille Packard Research Fellow}
\altaffiltext{2}{Kavli Institute for Particle Astrophysics and Cosmology, Stanford University, P.O. Box 20450, MS29, Stanford, CA 94309, USA}
\altaffiltext{4}{Institut dÕAstrophysique de Paris, UMR7095 CNRS; Univ. Pierre et Marie Curie, 98bis Bvd Arago, F-75014 Paris, France}
\altaffiltext{5}{Department of Physics and Astronomy, University of Utah, Salt Lake City, UT 84112, USA}
\altaffiltext{6}{Kapteyn Astronomical Institute, University of Groningen, P.O. Box 800, 9700AV Groningen, The Netherlands}
\altaffiltext{7}{Jet Propulsion Laboratory, California Institute of Technology, 4800 Oak Grove Drive, M/S 169-237, Pasadena, CA 91109, USA}


\begin{abstract}

We exploit the strong lensing effect to explore the properties of
intrinsically faint and compact galaxies at intermediate redshift ($\zs \simeq 0.4-0.8$) at
the highest possible resolution at optical wavelengths. Our sample
consists of 46 strongly-lensed emission line galaxies discovered by
the Sloan Lens ACS (SLACS) Survey. The galaxies have been imaged at
high resolution with \hst\ in three bands (\hstV, \hstI\ and \hstH),
allowing us to infer their size, luminosity, and stellar mass using
stellar population synthesis models. Lens modeling is performed using
a new fast and robust code, \klens, which we test extensively on real
and synthetic non-lensed galaxies, and also on simulated galaxies
multiply-imaged by SLACS-like galaxy-scale lenses. Our tests show that
our measurements of galaxy size, flux, and \sersic\  index are robust and
accurate, even for objects intrinsically smaller than the \hst\ point
spread function. The median magnification is 8.8, with a long tail
that extends to magnifications above 40. Modeling the SLACS sources
reveals a population of galaxies with colors and \sersic\  indices
(median $n\sim1$) consistent with the galaxies detected
with \hst\ in the GEMS and HUDF surveys, but that are (typically) $\sim2$
magnitudes fainter and $\sim5$ times smaller in apparent size than GEMS and 
$\sim4$ magnitudes brighter than but similar in size to HUDF.  The size-stellar
mass and size-luminosity relations for the SLACS sources are offset 
to smaller sizes with respect to both comparison samples. The
closest analog are ultracompact emission line galaxies identified by
\hst\ grism surveys.  The lowest mass galaxies in our sample are
comparable to the brightest Milky Way satellites in stellar mass
($10^7\msun$) and have well-determined half light radii of 
$0\farcs05$ ($\approx 0.3$ kpc).
\end{abstract}


\section{Introduction}

Understanding how galaxies were formed and how they evolve into those
we see today is an important cosmological question. In
hierarchical galaxy formation, gas condenses and cools
within a dark matter halo. To form disk galaxies, tidal torques impart
angular momentum to the dark matter halo and associated baryons;
angular momentum is conserved as the disk galaxy forms within the halo
\citep{F+E80}.  The most massive late-type galaxies are predicted to
have formed from mergers of smaller progenitors. However, such models
are complicated by the details of star formation, feedback processes,
cluster interactions, and effects of the bulge \citep[e.g.][]{MMW98a}.

Models for galaxy formation and evolution predict certain relations
between the basic physical properties of galaxies (i.e. luminosity,
size and mass); quantifying these relations can help test the
standard paradigm of galaxy formation and place limits on future models.  The
size-magnitude (or luminosity) and size-mass relations are
well-studied locally \citep[e.g.][]{She++03,Dri++05}.  The relations
for early and late-type galaxies (typically defined as having \sersic\ 
index $n>2.5$ or $<2.5$) are found to diverge.

\citet{She++03} looked at the mass-size and size-magnitude relations
for galaxies in the Sloan Digital Sky Survey (SDSS). A characteristic
mass of $M_*=10^{10.6}\msun$ delineates two regimes for the SDSS
size-mass relation: above this mass, the relation is steeper and
tighter ($\Reff \rm (kpc) \propto M_*^{0.39}$, $\sigma_{ln
R}=0.34$~dex) than for less massive galaxies (with $\Reff \rm (kpc)
\propto M_*^{0.14}$, $\sigma_{ln R}=0.47$~dex).  The behavior of the
size-magnitude relation is similar.

Surveys of intermediate ($0.1<z<1$) and high ($z>1$) redshift galaxies
have attempted to extend studies of the size-mass and size-magnitude
relations \citep[e.g.,][]{Fer++04, Bar++05, Mci++05, Tru++06,
Mel++07b}.  GEMS, an \hst\ survey, allowed \citet[]{Bar++05} to study
the magnitude-size and mass-size relations of late-type galaxies out
to $z\sim1$. They found a dimming of $\sim1$ mag from this redshift to
$z=0$ in the rest-frame V-band, but noted that the mass-size relation stays constant.
Using the Hubble Deep Field South, \citet[]{Tru++06} extended the SDSS
and GEMS work out to $z\sim2.5$, for the most luminous and massive
galaxies. The authors find that for low \sersic\  indices, galaxies at a
given luminosity were $\sim3.0\pm0.5$ times smaller at $z\sim2.5$,
while galaxies at a given mass were $\sim2.0\pm0.5$ times smaller.


\citet[]{Mel++07b} inferred the size-magnitude relation for blue galaxies in the
Great Observatories Origins Deep Survey \citep[GOODS]{Gia++04}.  These
authors found $\sim1.6$ magnitudes of dimming in the B-band since
$z\sim1$ for large and intermediate-sized galaxies ($\Reff > 3$~kpc),
in agreement with GEMS. Small galaxies, on the other hand, were found
to have dimmed significantly more, by some $2.55\pm0.38$ magnitudes in
B; this significant evolution is hypothesized to be the result of the
fading of the starburst galaxies rather than strong evolution of the
entire small galaxy sample.

These studies are limited by the resolution and completeness of the
\hst\ surveys: for GEMS, these limits correspond to galaxies with
$M_*>10^{10} \msun$ and $M_V<-20$, as determined by the highest
redshift bin ($z\sim1$).  Meanwhile, \citet[]{Tru++06} could only look
at galaxies with $L_V > 3.4\times10^{10}h^{-2} \lsun$ and stellar mass
$M_*> 3\times10^{10}h^{-2}\msun$.


One method of reaching to lower luminosities is to use exceptionally long exposure times: the Hubble Ultra Deep Field \citep[HUDF;][]{Bec++06} is the prime example of this.  Much of the work on the HUDF has been centered around high redshift galaxies ($z > 1$), but several authors have looked at the low redshift sample.  This sample is primarily comprised of faint objects, since the field was directed away from nearby bright galaxies.  \citet[]{Coe++06} observe large numbers of faint blue galaxies with magnitudes as low as $M_B = -14$ at $z=0.7$ -- believed to be young starburst galaxies -- which peak at $z\sim0.67$.  Two additional spectral energy distribution (SED) templates with a steep rise towards bluer wavelengths were added to accommodate these objects.  \citet[]{C&D07} look at the size-luminosity relation for HUDF galaxies with $0.2 < z < 1.15$. The best-fit evolution scenario is one with a brightening of 0.9 mag and a 5\% decrease in size from $z\sim0.1$ to $z = 0.675$.  This is consistent with the $~1$ mag arcsec$^{-2}$ of dimming since $z\sim1$ found in other surveys.  

Gravitational lensing is 
another method of extending surveys to
potentially smaller, fainter, and less massive galaxies and does
not rely on extremely deep imaging.  In strong
lensing, a massive foreground galaxy deflects the light from a
background object, resulting in multiple images of the source being
seen.  The source, in addition to being distorted, is typically
magnified by a factor of $\sim10$.  This phenomenon allows the study
of objects smaller than otherwise possible: the tiny source galaxy of
gravitational lens J0737$+$3216 was studied by
\citet[]{Mar++07}, while \citet{Sta++08} and \citet{Swi++09}, for
example, have used lenses to carry out detailed studies of high
redshift galaxies.  

In this paper, we study a sample of gravitationally-lensed galaxies,
selected from the Sloan Lens ACS Survey \citep[SLACS;][hereafter
\paperV\ and \paperIX]{Bol++08a,Aug++09}.  These objects
were previously modeled in the F814W filter; here, we perform
multi-filter modeling, which allows us to reconstruct the source
galaxy spectral energy distribution (SED), and hence infer its stellar
mass.  Our aim is to investigate the size-mass and size-magnitude
relations for these galaxies, and thus explore the potential of
gravitationally lensed galaxies to further the study of galaxy
formation and evolution.

This work is organized as follows. We introduce our lens sample in
\secref{sec:slacsintro}, then give our multi-filter source and lens
models for the SLACS sources in \secref{sec:slacs}.  We discuss the
properties of the lensed sources and make a comparison to the GEMS
and HUDF samples in \secref{sec:slacs:cfgems}.  In \secref{sec:elgs}, we compare
our galaxies to those found in recent emission line surveys. 
We discuss and summarize our results in \secref{sec:conc}.
A discussion of our new lens-modeling code, designed to perform fast
and robust lens modeling on large numbers of images, is reserved for
the Appendix; we also present tests on simulated gravitational lenses
and on both real and simulated non-lensed galaxies here.

Throughout, we assume a flat $\Lambda$CDM cosmology with $h=H_0/(100$~km~s$^{-1})=0.7$ and
$\Omega_{\rm m}=0.3$.  Apparent magnitudes are given in the AB system
unless otherwise stated, while Johnson V and B-band absolute
magnitudes are given in the Vega system.  All sizes are effective
radii, and we follow \citet{Pen++02} and \citet{Bar++05} and define
them on the major axis: for an elliptically-symmetric surface
brightness distribution, the elliptical isophote containing half the
total flux has major axis~$\Reff$ and subtends area $\pi q\Reff^2$
where $q$ is the ellipse axis ratio.


\section{The SLACS lens sample}
\label{sec:slacsintro}

Here we introduce our sample, a set of galaxies at redshift $\zs
\simeq 0.4-0.8$ being multiply-imaged by massive galaxies lying at
$\zd\simeq0.2$.


\subsection{Sample selection}
\label{sec:slacsintro:sample}

We use a subset of lenses from the SLACS survey \citep[]{Bol++08a} as
our sample.  We select those lenses which were classified as
``definitely a lens'' and were imaged in the \hstV, \hstI, and \hstH\
bands. The instruments and filters these refer to are as follows: for
the \hstV-band, ACS/WFC F555W or WFPC2 F606W, for the \hstI-band,
ACS/WFC or WFPC2 F814W, and for the \hstH-band, NICMOS/NIC2 F160W. For
the majority of the observations a full orbit's exposure ($t_{\rm exp}
\sim 2000$~sec) is available, except for 15 lenses that have only \hst\
Snapshot images in ACS/WFC F814W ($t_{\rm exp} = 420$~sec). For
details on the observations and data analysis, as well as full object coordinates,
see \paperIX. All images
were drizzled onto a pixel scale of $0\farcs05$. This multi-filter
sample comprises 46 lensed galaxies.


\subsection{Subtraction of lens galaxy light}
\label{sec:slacsintro:subtr}

When performing lens modeling, we attempt to fit the often-faint
lensed images, which can be hidden by or confused with light from the
lens (foreground) galaxy; thus, it is necessary to remove the light of
the lens galaxy prior to modeling.  As the SLACS survey preferentially
selects bright lens galaxies this is particularly important for our
work.  We use the radial B-spline technique, first introduced by
\citet[][in the appendix]{Bol++06}, for this purpose.  We refer the
reader to \paperV\ for a full discussion of this method, but provide
a brief summary here.  First, zero-weight pixels, neighboring objects
and potential source galaxy features are masked. Second, the data is
fit using only monopole, dipole and quadrupole terms in the angular
structure.  The lensed features are masked based on the residual image
from this initial fit; the image is then re-fitted using higher order
multipole terms as necessary. \citet{Mar++07} estimated that the
systematic errors in the source size and brightness due to the
subtraction of lens galaxy light by this method are approximately 2\%
($0.01$~kpc for a $0.6$~kpc source at $\zs$=0.6) and $0.10$~magnitudes
respectively. This is an important but usually not dominant source of
systematic error. Similarly, subtraction of the lens galaxy may
introduce a systematic uncertainty on the \sersic\  index, similar to
those introduced by improper sky subtraction; based on the detailed
work by \citet{Mar++07}, we estimate this systematic uncertainty to be
of order 0.1-0.2. This is larger than random errors, but does not effect 
the results of this paper.


\section{The SLACS sample: mass models and measurements of source observables}
\label{sec:slacs}

In the Appendix, we demonstrate our ability to measure galaxy sizes and magnitudes
through galaxy-scale gravitational lenses with \klens; here we use \klens\  to measure
the size, magnitude, and \sersic\  index of the SLACS sources
themselves.  We refer the interested reader to \paperV\ and \paperIX\ for images of the
SLACS lenses prior to lens galaxy subtraction and the reconstructed source planes
in the F814W filter. 


\begin{figure*}[!ht]
\centering\includegraphics[width=0.95\linewidth]{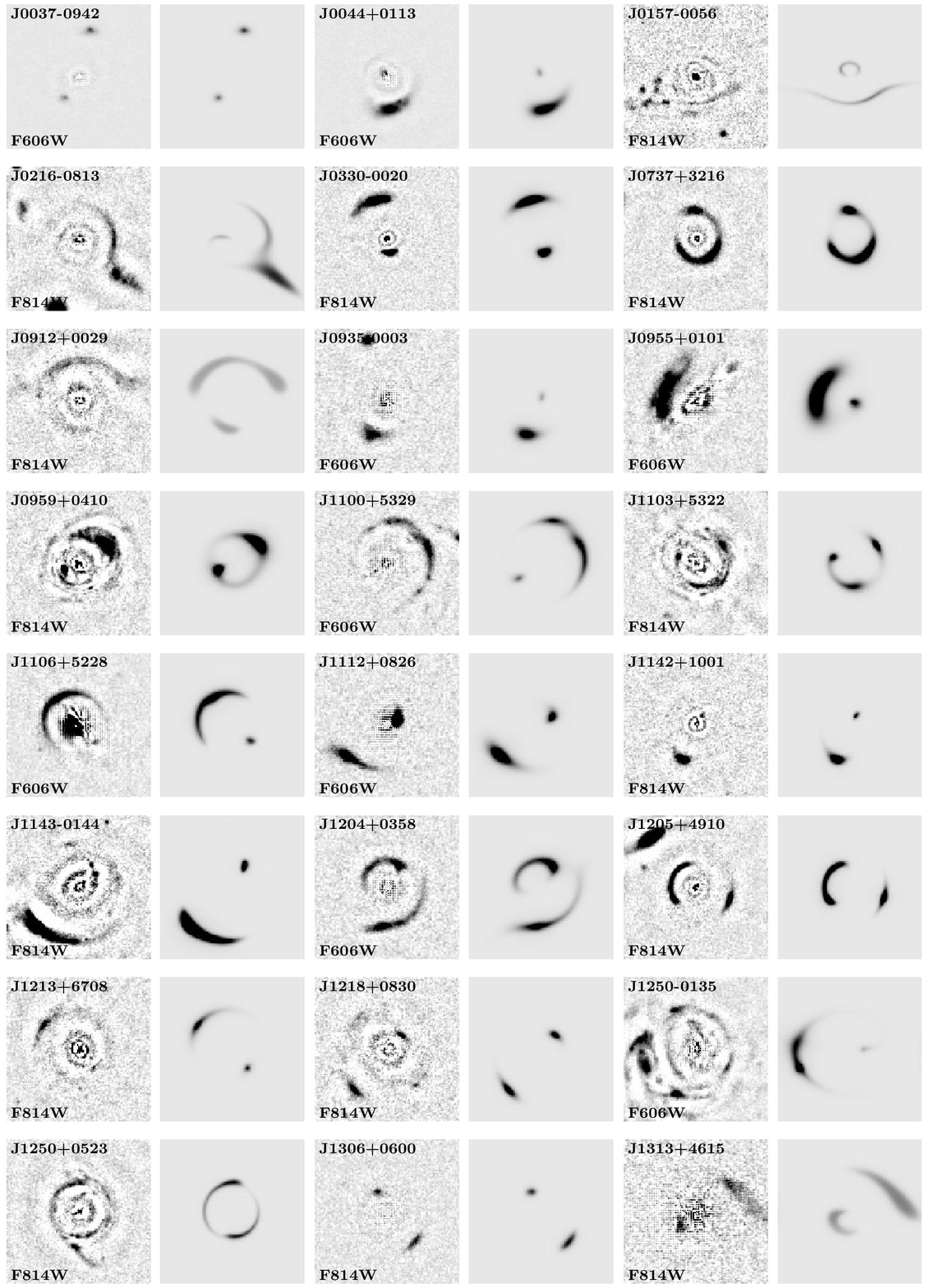}
\caption{Lens-subtracted \hst\ images with \klens\  model-predicted arcs. In each
case the image in the band with the highest signal/noise ratio is shown --
this is the image that was used when fitting the lens model.  Images are $6\times6''$.
\label{fig:montage}}
\end{figure*}

\begin{figure*}[!ht]
\centering\includegraphics[width=0.95\linewidth]{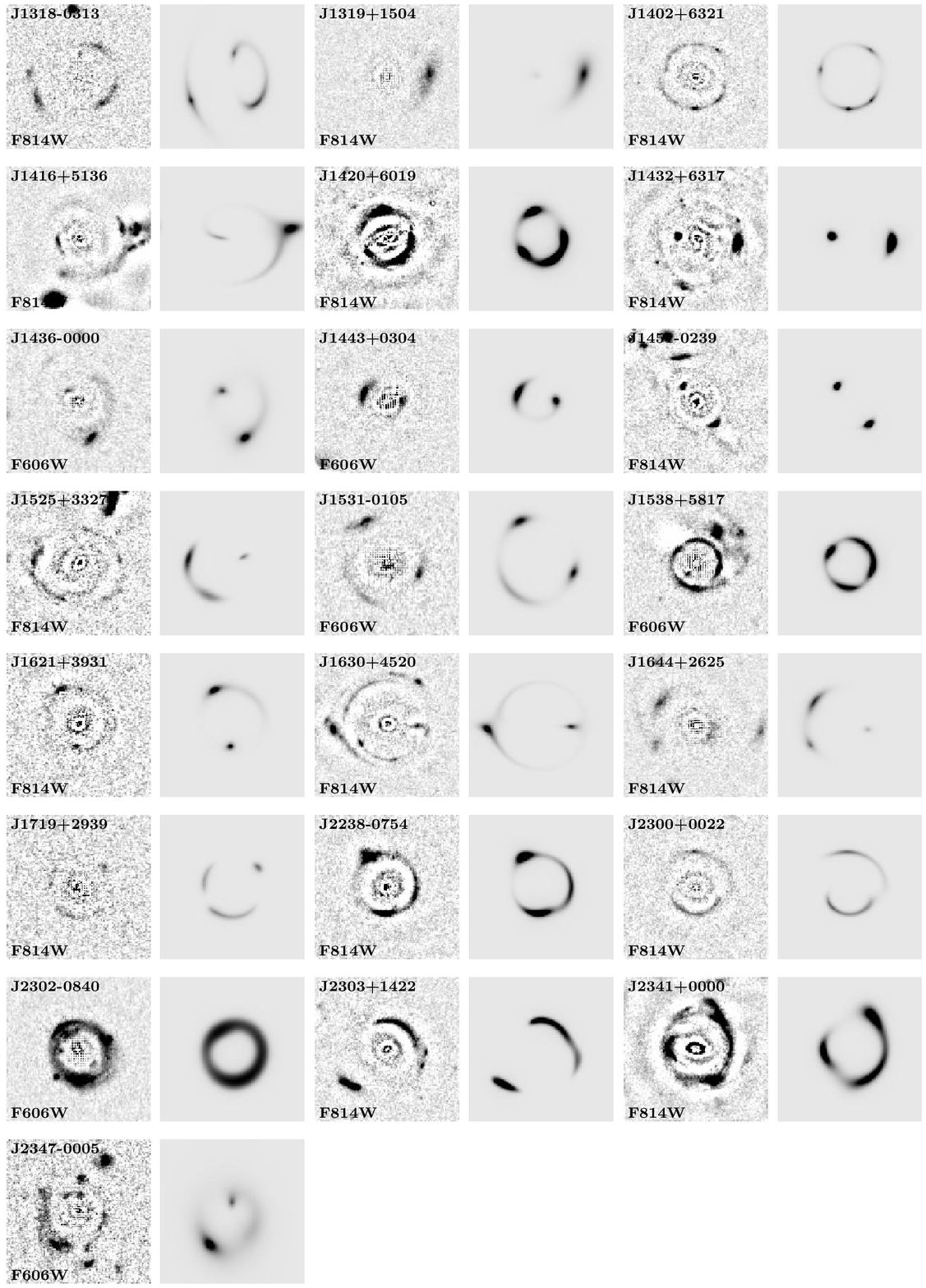}
\figurenum{\ref{fig:montage}}
\caption{{\it continued.}}
\end{figure*}


\subsection{Models of lens and source galaxies}
\label{sec:slacs:models}

\input{tab1.tex}

We have modeled the SLACS lenses in three filters; the data and model
lens planes for the primary modeling filter are shown in
Figure~\ref{fig:montage}.  In order to obtain robust and
easy-to-interpret constraints on the source galaxies, we make three
fairly standard assumptions, for which we give both {\it a priori}
justification and {\it a posteriori} validation:

{\it Assumption 1:} As in the testing program described in the
Appendix, we use only one source galaxy, even though in a
few cases the source is evidently more complicated
\citep[see][]{Bol++08a}.  Without this assumption, we would be unable
to interpret the size and stellar mass of the galaxy. The alternative
would be to use a more complex source model, perhaps defined on a
pixelated grid \citep[\eg][]{W+D03,T+K04,B+L06,Suy++06,V+K09}, and
then derive the size and magnitude of the source from this complex
model. However, imposing a simple, one-component \sersic\  model is
standard in the study of faint galaxies, even though they may have
irregular morphologies or multiple knots of star formation
\citep[e.g.][]{Fer++04}. If we are to compare our sample with others
in the literature, we need to derive the same parameters as were
studied by previous authors. A potential concern is that
oversimplification in the source surface brightness distribution may
lead to small biases in the mass model. However, as we discuss in
\secref{sec:slacs:cfpaperV}, comparison of our mass model parameters
with those inferred using complex multicomponent source distributions
shows that, at least for the SLACS sources, this is a negligible source
of error. In conclusion, considering that using a simple analytic
source model greatly speeds up computation time ($\sim 15$ minutes in
our approach compared with several hours for a typical pixel-based
source reconstruction, with a standard desktop CPU), we adopt this
procedure.

{\it Assumption 2:} We require that the mass model be identical across
the different filters, in order not to bias our source reconstruction.
Although ideally this would be avoided naturally because there is only
one deflector potential, lens galaxy subtraction and differences in
contrast and signal-to-noise ratio (S/N) can result in small
differences, which we quantify as follows.  Using the \hstI\  band as
reference, we found the difference in $\sigmaSIE$ to be, on average,
0.6~km~s$^{-1}$ with an r.m.s. scatter of 13.5~km~s$^{-1}$ in the \hstH\  band, and
$-2.0$~km~s$^{-1}$ with an r.m.s. scatter of 11.5~km~s$^{-1}$ in the \hstV\  band.
Likewise, the difference in mass distribution axis ratio $q$ was, on
average, $-0.007$ with an r.m.s. scatter of 0.15 in the \hstH\ band and
$-0.03$ with an r.m.s.  scatter of 0.12 in the \hstV\ band.  Two or
three outliers (defined as having $|\Delta \sigmaSIE|>50$~km~s$^{-1}$ or $|\Delta
q|>0.6$) were excluded from each of these calculations.  Fixing the
mass model to that of the highest S/N reconstruction
ensures that we do not introduce any unnecessary extra scatter in the
properties of the source.

{\it Assumption 3:} We fix the source morphology parameters (position
angle, inclination, size and \sersic\  index) to the best-fit values from
the filter with highest S/N. 
(For two-thirds of galaxies this filter is
F814W; for the remainder, only snapshot images were available in 
F814W and the filter with the highest S/N is F606W.  Those objects with
only snapshot images in F814W are indicated in Table \ref{tab:obs}.)  
This is necessary because the
S/N sometimes differs significantly between filters, which can cause
low-surface brightness features to be missed and result in overly
small sizes and \sersic\  indices, which in turn affects the inferred
magnitude. This procedure, analogous to the SDSS model magnitudes
\citep[][and references therein]{SDSSDR7}, is effectively equivalent
to measuring colors within fixed aperture, and is widely adopted in
order to obtain robust colors when the S/N varies significantly
between filters.

Inferred, unlensed angular sizes and apparent magnitudes for the
source galaxies are given in Table~\ref{tab:obs}, along with their
lens and source redshifts, and a flag indicating whether the
\hstV\ band magnitude refers to WFPC2/F606W or ACS/F555W. We also give
the total magnification $\mu$ of each system. This quantity varies
widely, between 2.4 and 44.3;
the median magnification is $8.8^{+6.7}_{-5.1}$, where the error bars
correspond to the 16 and 84$^{\rm th}$ percentiles.
The lens galaxy properties -- mass axis ratio, inclination, and the 
velocity dispersion --  are not presented here, as our concern is with the 
source galaxies.  For a thorough treatment of the SLACS lens galaxies, 
we refer the reader to \paperV\ and \paperIX.


\subsection{Testing mass models: comparison to previous work}
\label{sec:slacs:cfpaperV}

\begin{figure}
\centering\includegraphics[width=0.9\linewidth]{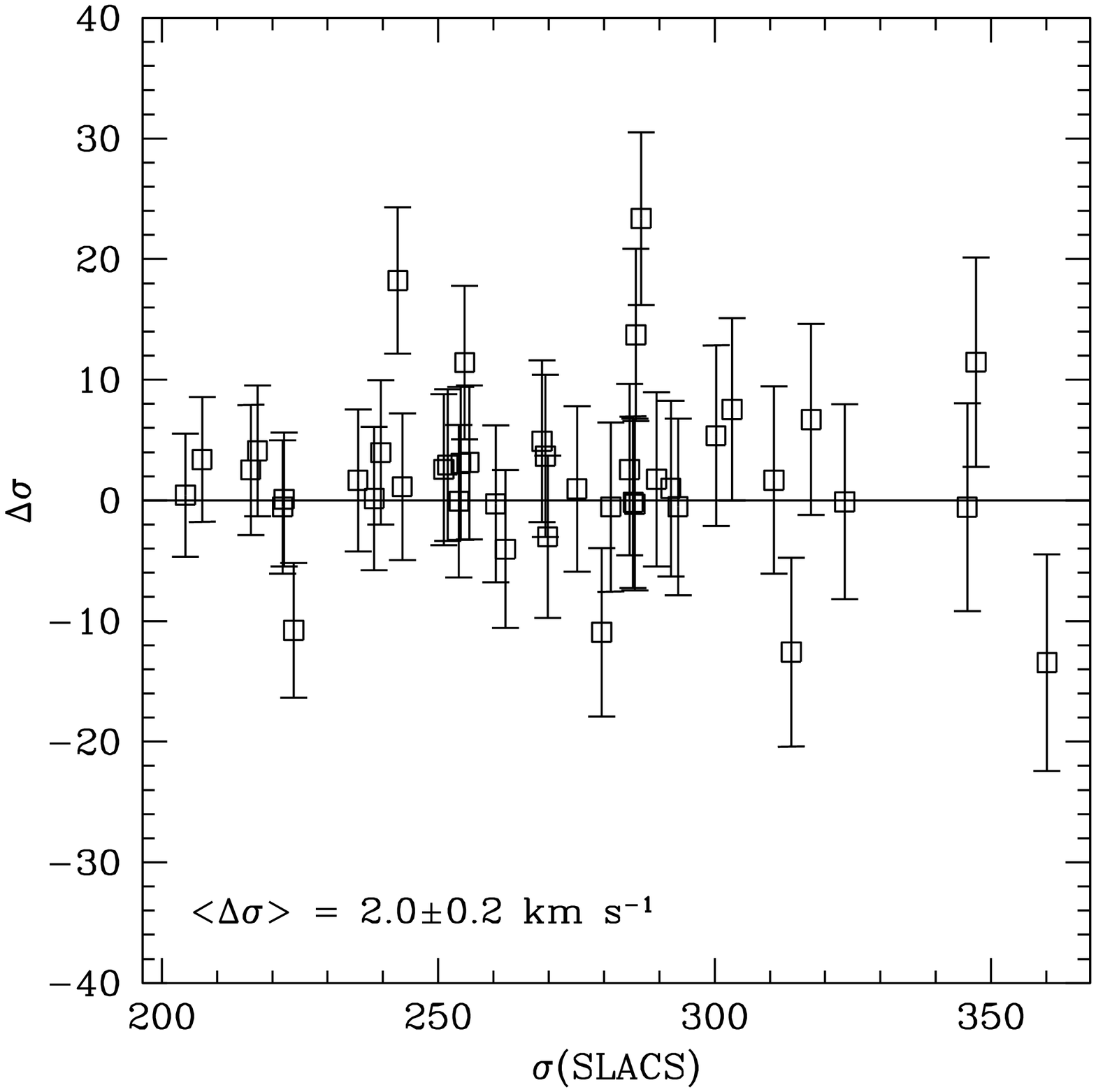}
\caption{Comparison of best-fit \klens\  velocity dispersions to the \paperV\ or
\paperIX\ value. SLACS values are on the $x$-axis and the difference between
\klens\  and SLACS is shown on the vertical axis. The average difference is
$2.0\pm0.2$ km s$^{-1}$, and using the r.m.s. scatter we estimate the error on
$\sigma_{SIE}$ to be 1.78\%. The error bars shown are total errors (2.5\%) on
the difference.  \label{fig:compsig}}
\end{figure}

The F814W images of the lenses in our sample have been modeled by
\citet[]{Bol++08a} and \citet[]{Aug++09} using multiple objects in the source
plane.  This gives us the opportunity to test \klens ' ability to model real
lenses and estimate systematic (modeling) errors. Although we cannot compare
source models (given the different definitions), we are able to look at the
effects of using only one source object on inferred mass model parameters. 
The lens mass model is described by the mass axis ratio~($q$), the inclination,
the velocity dispersion of the best-fitting singular isothermal ellipsoid 
($\sigmaSIE$) and the mass centroid.
We compare mass axis ratios and velocity dispersions and find that they agree to 
within levels expected from systematic errors: the average offset between our fits 
and those from \paperV\ is~$-0.015\pm0.002$ for $q$ and~$2.0\pm0.2$ km s$^{-1}$ for
$\sigmaSIE$.  The comparison for velocity dispersion is shown in Figure~\ref{fig:compsig}.

We can use the scatter of these relations to estimate total errors,
including systematics: we calculate~$0.07$ for the error on~$q$
and~$1.78\%$ for the error on~$\sigmaSIE$.  These are slightly larger
than the errors adopted in \paperV\ ($0.05$ on $q$ and $1.0$\% on
$\sigmaSIE$), as expected because our models are less flexible. As
discussed by \citet{Mar++07} and in the summary at the end of the
Appendix, if the potential is not perfectly described by a SIE
\citep[e.g.][]{Koo++09,Aug++10} then additional errors will apply to
the source galaxy properties. These are taken into account in the
analysis of the source population.

We note that one of the systems, J0737$+$3216, has a published source
size, magnitude and stellar mass~\citep{Mar++07}. Even with a
completely different lens modeling code, we find lens parameters,
source size and source \sersic\  index for this system that are
consistent with those in this previous work, providing further
confirmation of the robustness of our approach. However, we infer a
source that is 1.5 magnitudes fainter uniformly in all bands. This was
traced to an error in the modeling code used by~\citet{Mar++07}, which
summed the flux in twice-subsampled pixels rather than averaging it:
this factor of 4 in flux translates to a magnitude difference of
1.5. After correcting for this ``bug'' the magnitudes are also in
excellent agreement. The results here supercede those published in
\citet{Mar++07}.

These comparisons show that our models are reliable and give answers
consistent with those obtained by other methods once systematic errors
are taken into account.


\subsection{Rest-frame luminosity and stellar mass of the source galaxies}
\label{ssec:restmass}

We now use our multi-band photometry of the SLACS sources to infer
rest frame luminosity in the B and V (Johnson Vega) bands as well as
stellar masses (M$_*$). This analysis is based on our models of the 
unlensed source galaxies, where modeling has been carried out in 
three separate bands.  For this purpose we use the Bayesian code
developed by \citet{Aug++09} to fit stellar population synthesis (SPS)
models to our multi-band source galaxy photometry. The code computes,
for each galaxy, the likelihood of SPS models as a function of stellar
mass, age, metallicity, star formation history and dust content. In
combination with a prior on each of these parameters, the likelihood
then gives a posterior PDF for each model parameter. The same models
and posterior can also be used to generate self-consistent rest-frame
luminosities.  For this application we adopt SPS models by
\citet{B+C03} and uniform priors on the logarithm of stellar mass,
age, metallicity, and time scale of the exponential star formation
history, as is appropriate for the situation where we do not know the
order of magnitude of these quantities.

\input{tab2.tex}

As discussed by \citet[][and references therein]{Aug++09}, although
parameters such as age and metallicity are often degenerate, stellar
masses and luminosities in the range of wavelength covered by the data
can be derived quite accurately for a given IMF. Typical errors on the
transformations to rest frame luminosities are of order 0.05-0.1 mags,
while typical errors on stellar mass are of order 0.1-0.2 dex (see
Table~\ref{tab:inf} for details).  For simplicity we neglect the
impact of emission lines on broad band photometry, which is estimated
to be of order a few percent for the typical H$\alpha$ fluxes (the
strongest line in the wavelength range of interest) of a few
10$^{-16}$ erg s$^{-1}$ cm$^{-2}$.  Typically, no strong emission line
is present in the \hstH-band filter, which is providing the bulk of
the information for stellar mass estimate.  The main residual source
of uncertainty is the normalization of the initial mass function. In
this paper we adopt the \citet{Kro01} normalization of the initial
mass function (IMF), to facilitate comparisons with the GEMS
work~\citep{Bar++05}.  The uncertainty in the lens mass density profile
slope is an additional source of error \citep[]{Mar++07}.

As a sanity check we compared our estimated stellar masses with those
inferred by applying the ``standard'' recipe by \citet{Bel++03} to our
inferred colors.  For the same IMF the stellar masses agree very well,
with an average offset of $0.02\pm0.02$ dex (r.m.s. scatter 0.15
dex). Our synthetic rest frame photometry agrees well with that
inferred by standard ``K-correction'' procedures (for example, by
comparing with the V band magnitudes inferred via a Sbc template we
find an average offset of $0.06\pm0.06$ magnitudes). We conclude that
our stellar mass and synthetic photometry are robust and unbiased
within the errors.


\section{Properties of the source population}
\label{sec:slacs:cfgems}

In this Section we study the properties of the source population. To
put our galaxy population in context, we use two comparison samples of
non-lensed galaxies observed with \hst\ at comparable redshift, by
the GEMS and HUDF collaborations. After a brief description of the two samples
(\secref{ssec:gems}), we investigate the distribution of observed
properties, i.e. \sersic\  index, magnitude, and effective radius
(\secref{ssec:cobs}). In \secref{ssec:crest} we investigate the
bivariate distribution of rest frame quantities, i.e. V-band
magnitude, B-V color, effective radius, and stellar mass. Finally, in
\secref{sec:elgs} we compare the properties of the SLACS sources to
those of emission line galaxies selected by blind spectroscopic
surveys.


\subsection{GEMS-selected comparison sample}
\label{ssec:gems}

For the first comparison sample, we follow \citet[]{Bar++05} and select,
from the publicly-available GEMS catalog,\footnote{The GEMS catalog
used in this work can be obtained from
\texttt{http://mpia.de/GEMS/gems\_20090526.fits}, or from M.~Barden on
request.} galaxies with successful \galfit\  structural fits
(\texttt{GEMS\_FLAG} = 4) matched within 0\farcs5 of a COMBO-17 object
with successful photometric redshift estimate (\texttt{COMBO\_FLAG} =
3). We did not select galaxies based on their size or \sersic\  index
\citep[other than to reject objects with large size or \sersic\  index
uncertainties, or that reached the \galfit\  boundary conditions, as
did][]{Bar++05}, but we did reject objects not detected in both F606W
and F850LP filters. The resulting GEMS sample comprised $6999$
galaxies with measured absolute magnitudes and sizes. We adopt the
F850LP-measured sizes, since these differ from the rest-frame V-band
sizes by typically only 3\% \citep{Bar++05}. We then select galaxies in
the redshift range $0.4<z<0.8$ to approximately match the redshift
distribution of SLACS sources, resulting in $3369$ galaxies.  These
are shown as small black points or hatched blue histograms in 
Figs.~\ref{fig:incmag}-\ref{fig:sizemass}. We note
that this GEMS subsample represents the brighter part of the GEMS full
\hst\ catalog with surface photometry, owing to the shallower COMBO-17
multiband photometry used to determine photometric redshifts
(completeness limit of $F850LP\sim23.5$ vs. $F850LP\sim24.5$ for the
\hst\ catalog; Barden et al.\ 2005).

We computed stellar masses for all objects in the GEMS sample by
applying the recipe by \citet[][for a Kroupa IMF]{Bel++03} to the rest
frame photometry provided in the GEMS catalog. As discussed before,
for the SLACS sample the \citet{Aug++09} stellar masses are in
excellent agreement with the ones obtained using the recipe by
\citet{Bel++03}, ensuring that we are able to make this comparison.

\subsection{HUDF-selected comparison sample}
\label{ssec:hudf}

For the second comparison sample, we use the galaxy
catalog\footnote{The HUDF catalogs used in this work can be obtained
from the original paper or from D.~Coe at
\texttt{http://adcam.pha.jhu.edu/~coe/UDF/}.}  from the HUDF analyzed
by \citet[]{Coe++06}.  We selected galaxies in a manner similar to the
process used for GEMS.  Galaxies included in our sample are those: (1)
with good \galfit\ fits, (2) that are matched to \citet[]{BSS03} ACS
objects within 0\farcs5 (3) with successful Bayesian photometric
redshifts and (4) which are detected in $B$, $V$, $i'$, and $z'$.  Our
criterion for successful \galfit\ and redshift fits is that the
$\chi^2$ be within 4.5$\sigma$ of the mean.  We also restrict our
sample to galaxies in the redshift range $0.4<z<0.8$ to match the
redshifts of the GEMS and SLACS source samples, for a final tally of
841 galaxies.  These are shown as black points or hatched red
histograms in Figures~\ref{fig:incmag}--\ref{fig:sizemass}.

We use \texttt{kcorrect} from \citet[]{B&R07} to
compute rest-frame luminosities of the HUDF comparison sample.  We
then use the \citet[]{Bel++03} algorithm to calculate stellar masses,
as described in \secref{ssec:gems}.

\subsection{Distribution of observed properties}
\label{ssec:cobs}

\begin{figure*}
\centering
\begin{minipage}{0.48\linewidth}
\includegraphics[width=0.9\linewidth]{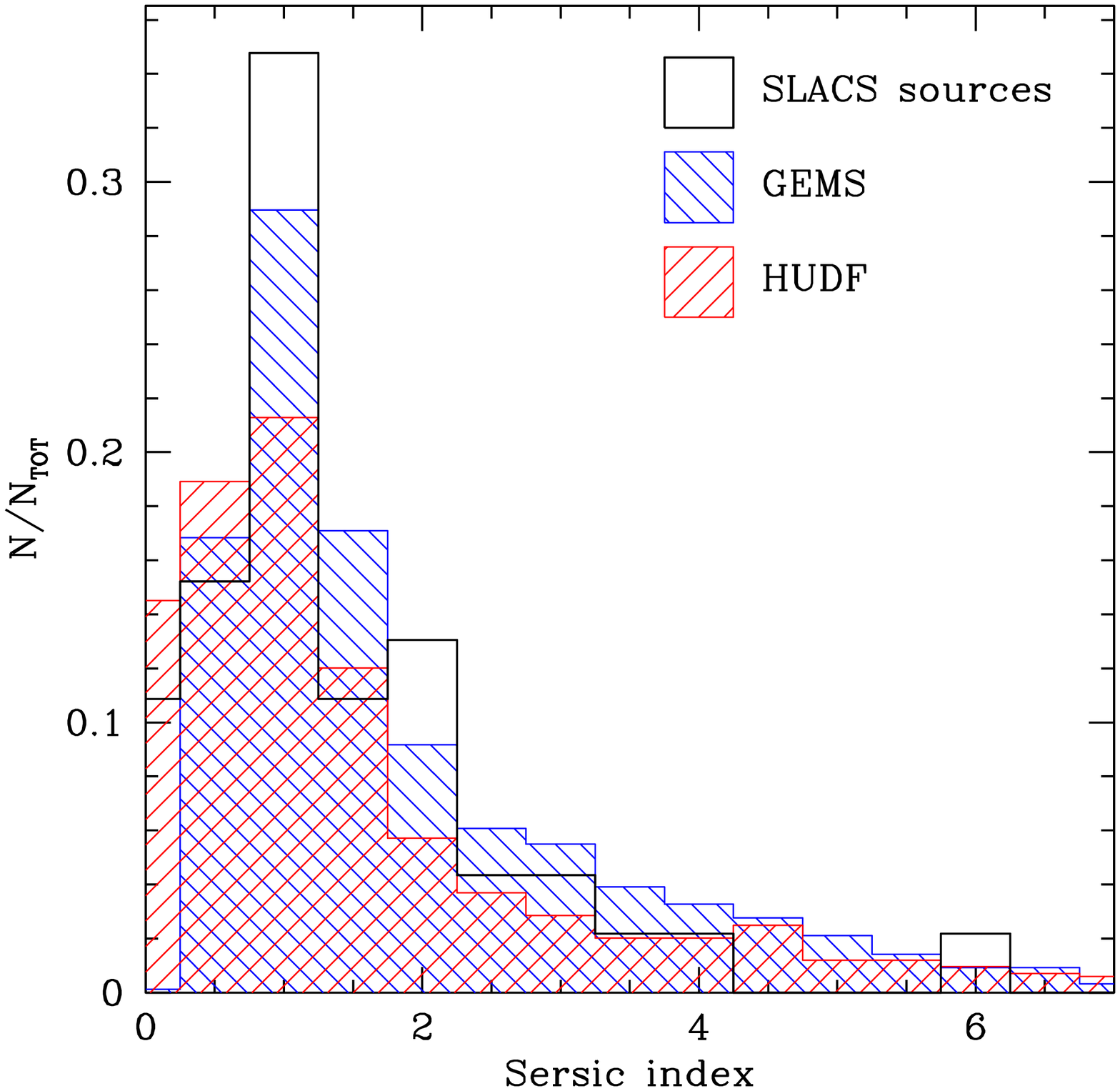}
\end{minipage}\hfill
\begin{minipage}{0.48\linewidth}
\includegraphics[width=0.9\linewidth]{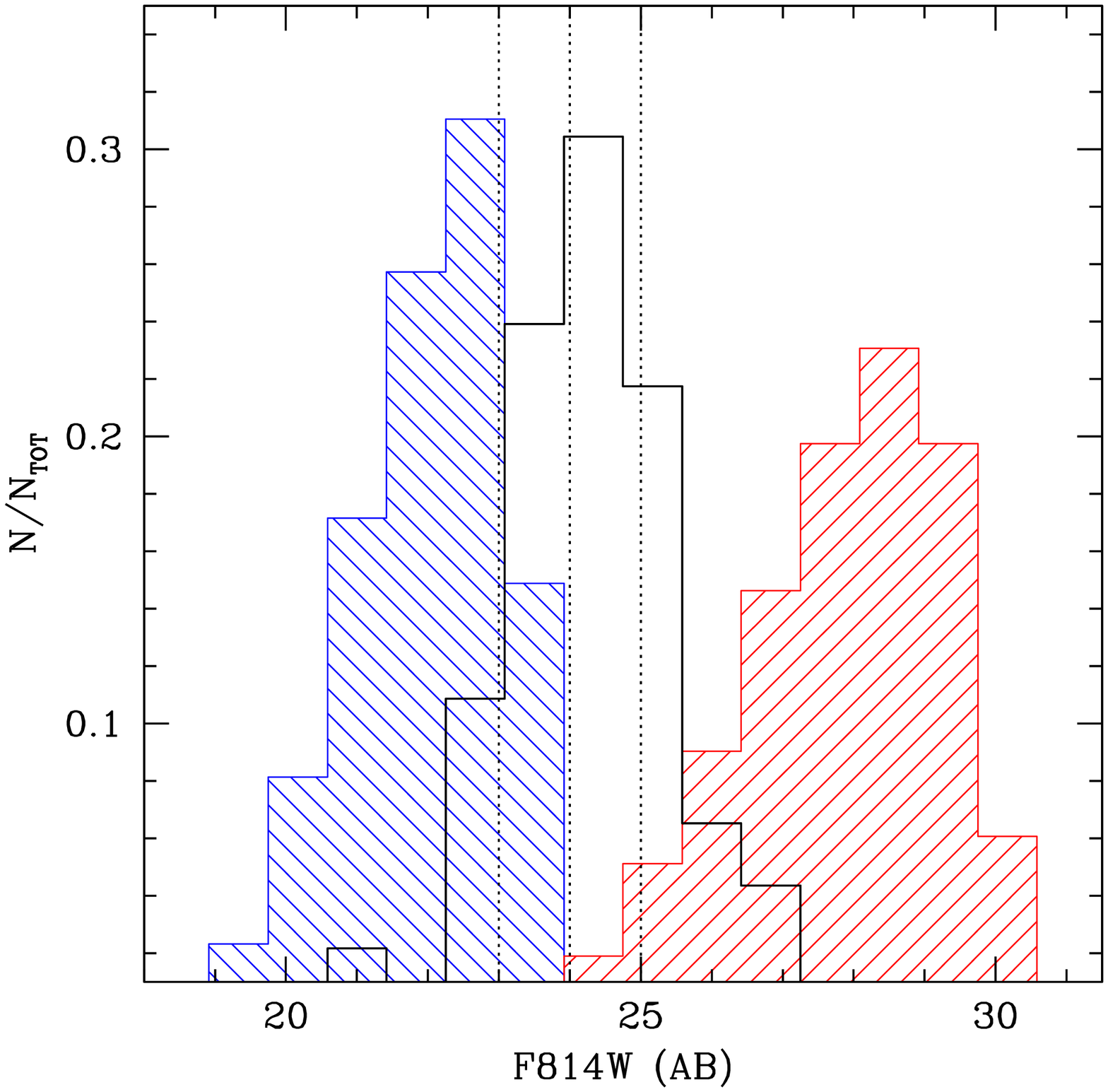}
\end{minipage}
\caption{Characteristics of the SLACS lensed sources.  The left panel
shows the distribution of \sersic\  indices for our galaxies (solid unfilled
histogram), for GEMS sample with $0.4<z<0.8$ (blue, hatched
histogram) and for the HUDF sample with $0.4<z<0.8$ (red, hatched
histogram). The right hand panel shows the distribution of source apparent
magnitudes, corrected for magnification. 
To compare with the GEMS and HUDF samples we have assumed that
F814W=F850LP=F775W (AB), for simplicity.  The vertical dashed lines indicate
the magnitudes in which KLENS was tested by simulations\label{fig:incmag}.}
\end{figure*}

We plot the distributions of source \sersic\  index and apparent F814W
source magnitude in Figure~\ref{fig:incmag}, showing both the SLACS
sources and the GEMS sample. The vertical dotted lines show the
magnitudes tested by our simulations, demonstrating that we are in a
regime where source properties can be reliably measured by \klens.
In this figure and throughout this work, the quoted source galaxy 
magnitudes are for the unlensed galaxy, i.e. corrected for 
magnification.


When considering apparent magnitudes, the SLACS sources 
lie squarely between the two regimes sampled by the GEMS
and HUDF surveys.  Most SLACS source galaxies fall in the range
$22<F814W<26$, peaking at magnitudes fainter than the completeness limit 
of the GEMS survey but brighter than most galaxies in the HUDF
sample.  As mentioned earlier, the drop-off of the GEMS catalog at $\sim23.5$ is
mostly due to the COMBO-17 completeness limit. However, a substantial
fraction of the SLACS sources are fainter than even the GEMS HST
completeness limit, suggesting that we are indeed exploring a
different population of intrinsically fainter objects.  The HUDF survey 
is much deeper than the SLACS survey (144 \hst\ orbits versus one in I band), so it is 
not surprising that these objects are fainter than the SLACS sources.
The average magnitudes for the SLACS, GEMS and HUDF surveys are 24.3, 22.0
and 27.5 respectively.

In contrast, the distribution of \sersic\ indices is essentially the
same for the three samples -- peaked at $n\sim1$ generally interpreted
as dominated by faint disks or compact galaxies -- even though the
magnitude ranges probed by the three surveys are very different.
The median for both SLACS and HUDF
is 1.1; for GEMS it is 1.3.

\begin{figure*}
\centering
\begin{minipage}{0.48\linewidth}
\includegraphics[width=0.9\linewidth]{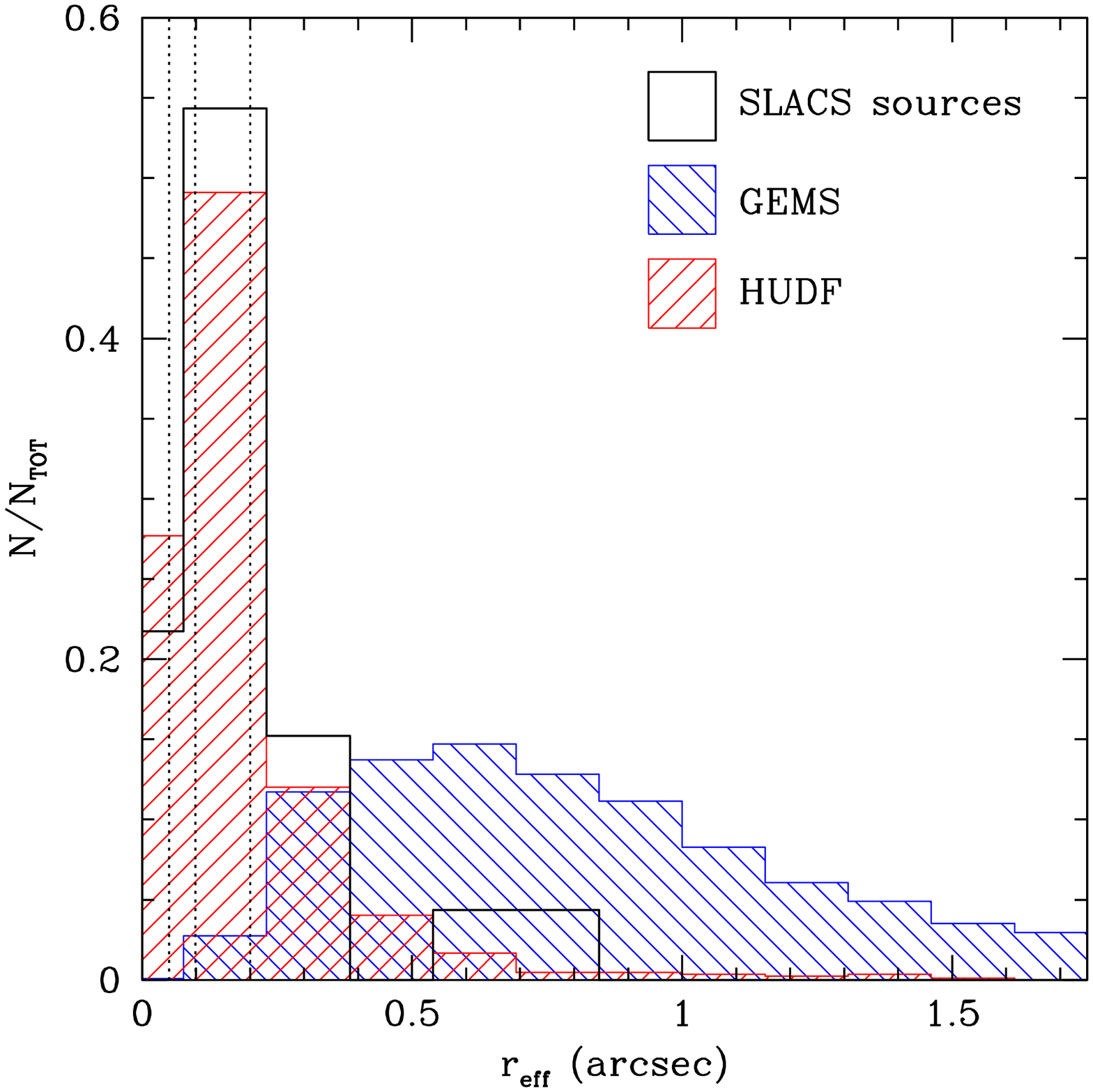}
\end{minipage}\hfill
\begin{minipage}{0.48\linewidth}
\includegraphics[width=0.9\linewidth]{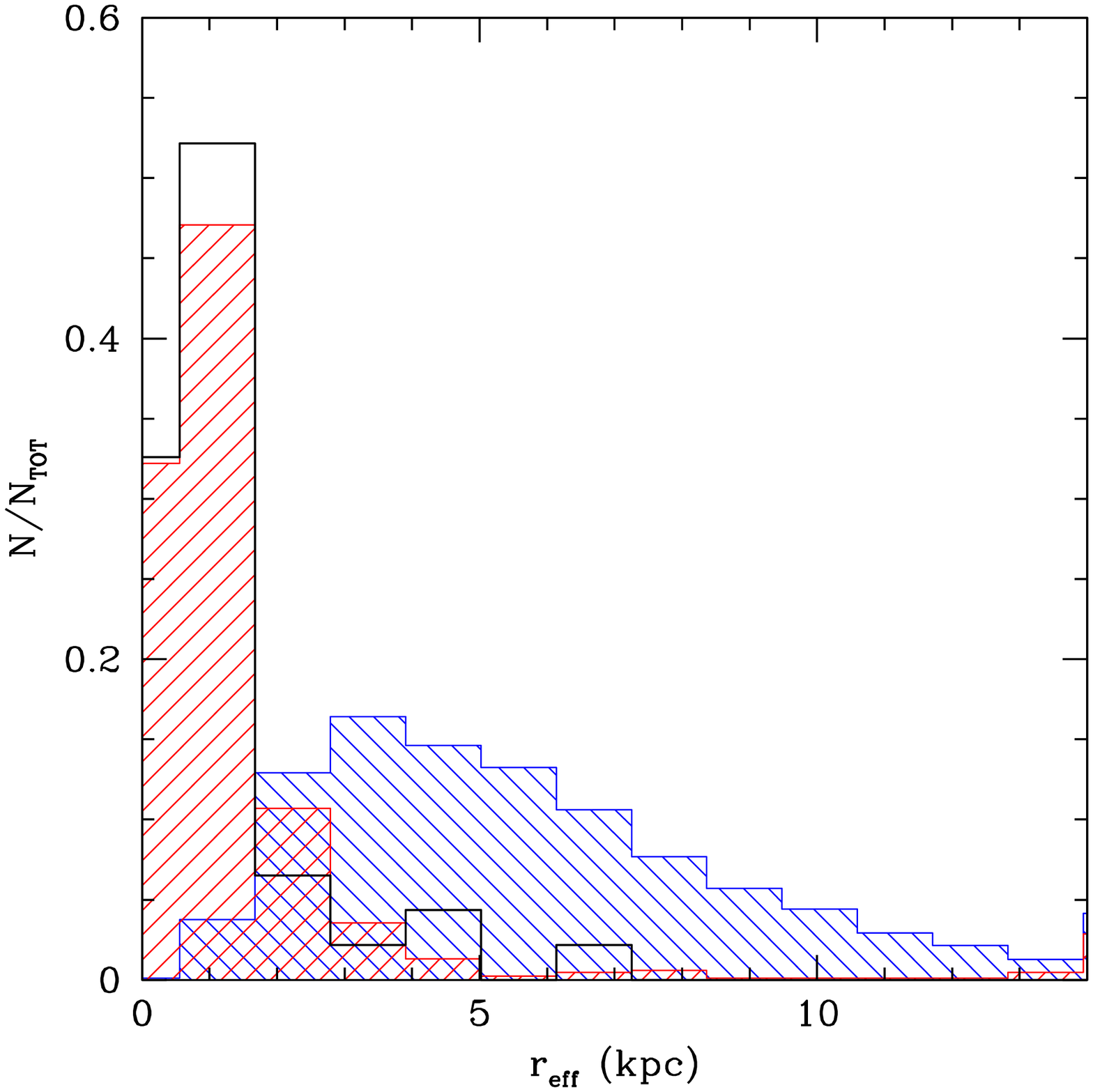}
\end{minipage}
\caption{Distribution of source effective radii for our galaxies. The
left-hand panel shows sizes in arcseconds; the vertical dotted lines
indicate the sizes tested by simulations. The right hand panel shows
sizes in kpc. The SLACS sources sample is shown as solid histograms;
the GEMS and HUDF comparison samples are shown as hatched
blue and red histograms respectively. \label{fig:reff}}
\end{figure*}

The difference between the SLACS sources and the GEMS galaxies is even
more pronounced in terms of the size distribution, shown in
Figure~\ref{fig:reff} with the same notation as
Figure~\ref{fig:incmag}.  The size distribution of GEMS sources is
much broader and peaks at $0\farcs78$, petering off below $0\farcs2$.
In contrast, Figure~\ref{fig:reff} highlights the similarity of the
SLACS and HUDF galaxies.  The overall size distributions for the two
samples is remarkably similar, though the HUDF sample extends to much
larger sizes, beyond the plotted region.  This long tail pushes the
average size up to $0\farcs35$ ($2.3$~kpc) for HUDF, compared to
$0\farcs19$ ($1.24$~kpc) for the SLACS sources.  The median sizes,
however, are more similar: $0\farcs12$ for HUDF and $0\farcs14$ for
SLACS, corresponding to $0.8$~kpc for both samples.

\begin{figure}
\centering\includegraphics[width=0.9\linewidth]{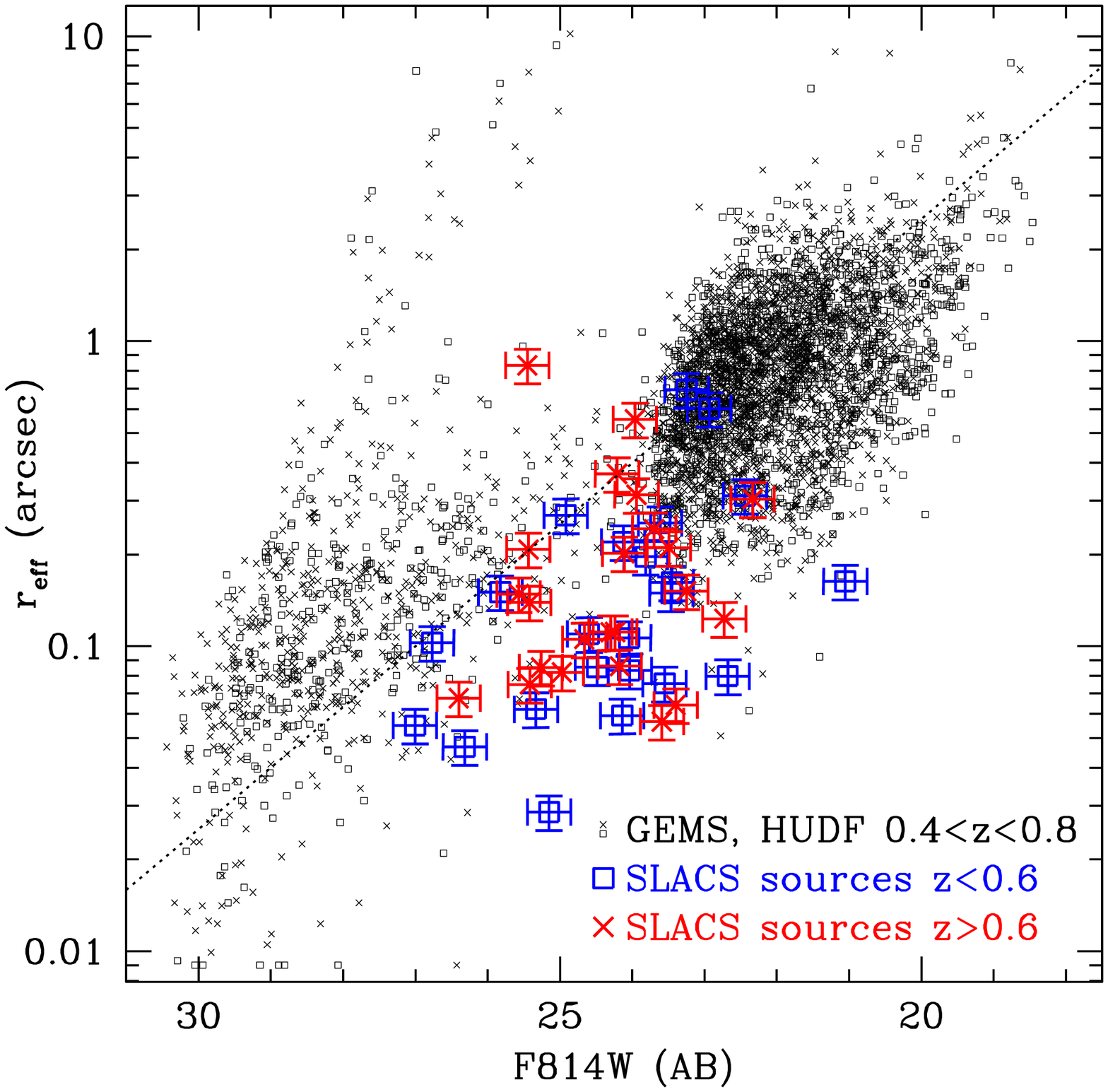}
\caption{Size-magnitude relation for SLACS source galaxies (large points
with error bars, with values corrected for magnification) 
and all GEMS and HUDF galaxies (small points). 
F814W=F850LP=F775W (AB) has been assumed in the comparison for
simplicity.  Crosses represent galaxies between $0.4<z<0.6$ 
and squares those between $0.6<z<0.8$.  
The diagonal line represents constant surface brightness
of 24 mag arcsec$^{-2}$ in F814W (AB), close to the completeness limit
of GEMS \label{fig:sizemag1} for a circular source.}
\end{figure}

The complementarity of the SLACS, GEMS and HUDF samples is further illustrated 
by Figure~\ref{fig:sizemag1} where sizes are plotted against apparent
magnitude. With one exception, SLACS sources have comparable surface
brightness to that of the GEMS sources, but extend much further
down in magnitude and size. A few of the brightest SLACS sources have
magnitudes comparable to those of the GEMS sources, but have sizes at
the compact end of the distribution. In contrast, the SLACS sources have
significantly higher surface brightness on average than the HUDF galaxies.

In Figure~\ref{fig:sizemag1}, crosses represent galaxies in the low redshift bin
($0.4<z<0.6$) while squares indicate the higher redshift galaxies ($0.6<z<0.8$).
The galaxies in all three samples are not strongly segregated by redshift, 
consistent with a broad distribution in intrinsic luminosity.

The comparison makes it apparent that the SLACS source population is
significantly different from the one studied by typical HST broad band
imaging surveys such as GEMS. These sources are most likely excluded
by \hst\ field surveys because they are too faint and/or too compact to
be detected and identified as galaxies (as opposed perhaps to
stars). Furthermore, even if detected, they are generally too faint
and compact to determine their surface brightness profile, size and
stellar mass.
The HUDF survey relies on its extremely large exposure times to
detect and model such sources. 
Conversely, SLACS identifies them as galaxies and measures their
redshifts via their emission lines (even though they are detected in
all three broad bands), and therefore their intrinsic faintness and
compactness does not represent an obstacle to detection, while
magnification helps in determining their structural parameters. In
fact, the distribution of SLACS sources and sizes is in agreement with
a generic feature of lensing survey. Owing to magnification bias,
faint and compact sources tend to dominate the source population when
their number density decreases sharply with luminosity and size
\citep[e.g.][and references therein]{Tre10}. We will return briefly to the
lensing selection function in the next section.


\subsection{Distribution of rest-frame quantities}
\label{ssec:crest}

We now explore the distribution in size, luminosity, color and
stellar mass for the SLACS source galaxies. We use two-dimensional
plots to examine bivariate distributions in the space of well known
correlations representing physical mechanisms: the size-luminosity
relation, the color-magnitude diagram, and the size-stellar mass
relation.  Again we use the GEMS and HUDF samples as a comparison.

\begin{figure}
\centering\includegraphics[width=0.9\linewidth]{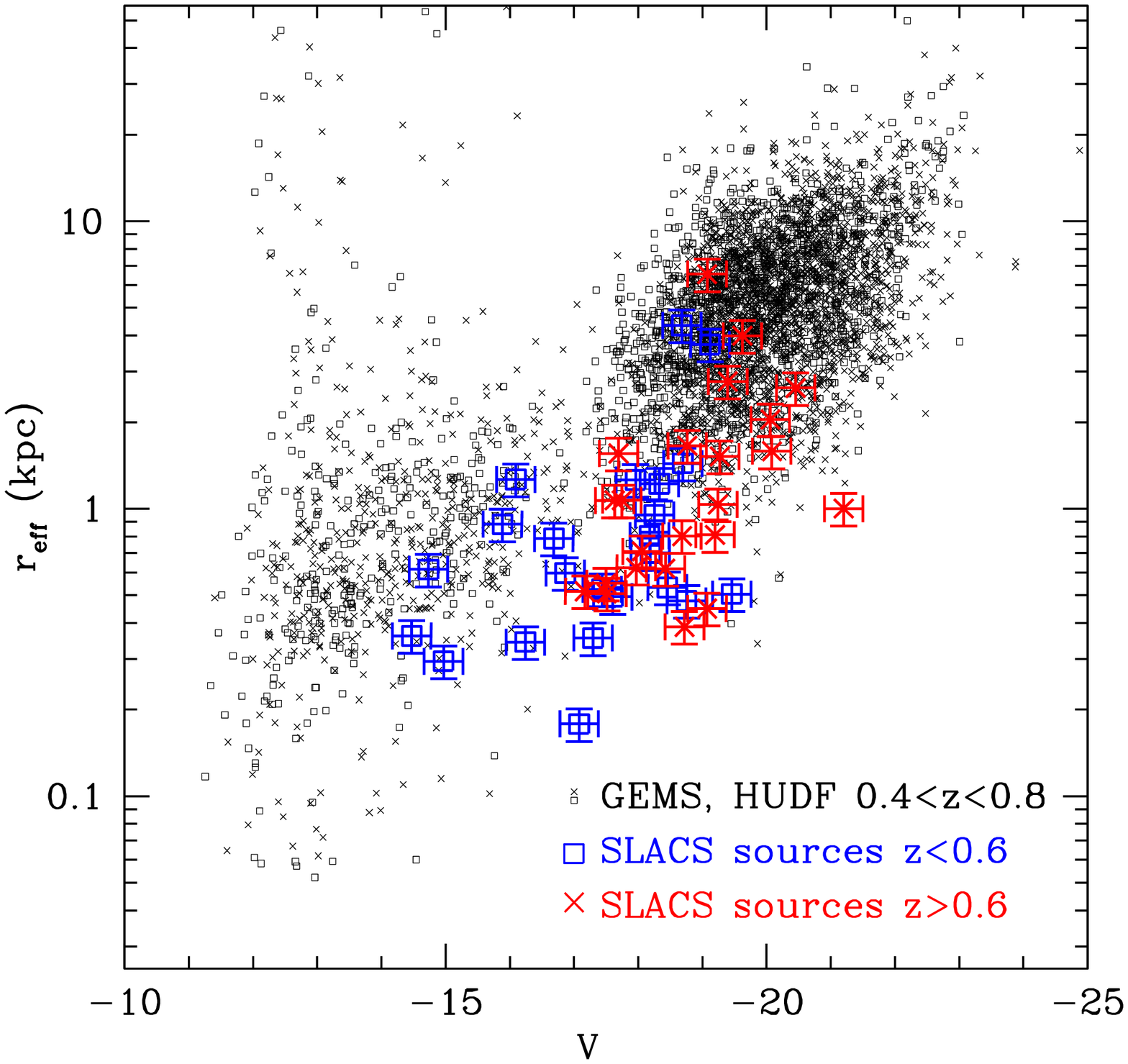}
\caption{Size-luminosity relation for SLACS source galaxies (large points
with error bars bars) and all GEMS and HUDF galaxies (small points).
Crosses represent galaxies between $0.4<z<0.6$ 
and squares those between $0.6<z<0.8$.  
\label{fig:sizemag}}
\end{figure}

We begin with the size-luminosity relation in Figure~\ref{fig:sizemag}. 
We fit a size-luminosity relation of: $\log_{10} \Reff \rm (kpc) =
-0.12 \times V - 2.25$. However, this result should be interpreted
with a care, keeping in mind the small number of objects in our sample
and the uncertain selection function.  A linear extrapolation of the
GEMS size-luminosity relation shows little offset between the GEMS and
HUDF galaxies, but the SLACS sources are offset towards smaller sizes
by $0.6$~kpc, or $\sim30\%$.
This offset may indicate that there is an extended tail of ultracompact 
galaxies of which the SLACS sources are a part, similar to the
population of compact emission line galaxies
\citep{Koo++95,Phi++97,Dro++05,Str++09}. 

\begin{figure}
\centering\includegraphics[width=0.9\linewidth]{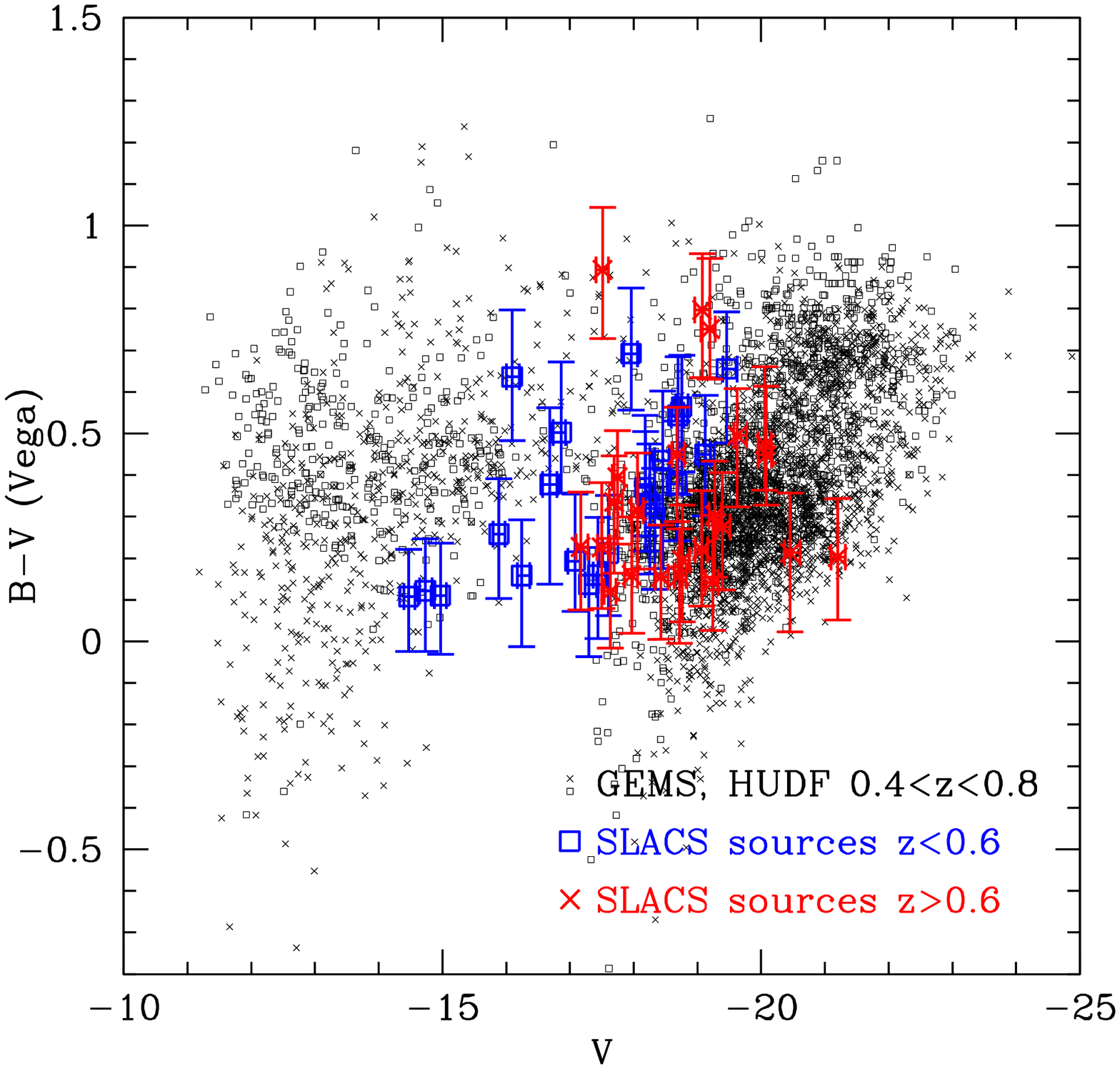}
\caption{Color magnitude relation for SLACS source galaxies (large points
with error bars bars) and all GEMS and HUDF galaxies (small points).
Crosses represent galaxies between $0.4<z<0.6$ 
and squares those between $0.6<z<0.8$.  
We use rest-frame B and V band magnitudes calculated as
described in \secref{ssec:restmass}. For our galaxies, we show the
rest-frame magnitudes calculated from our stellar mass fitting
code. \label{fig:colormag}}
\end{figure}

\begin{figure}
\centering\includegraphics[width=0.9\linewidth]{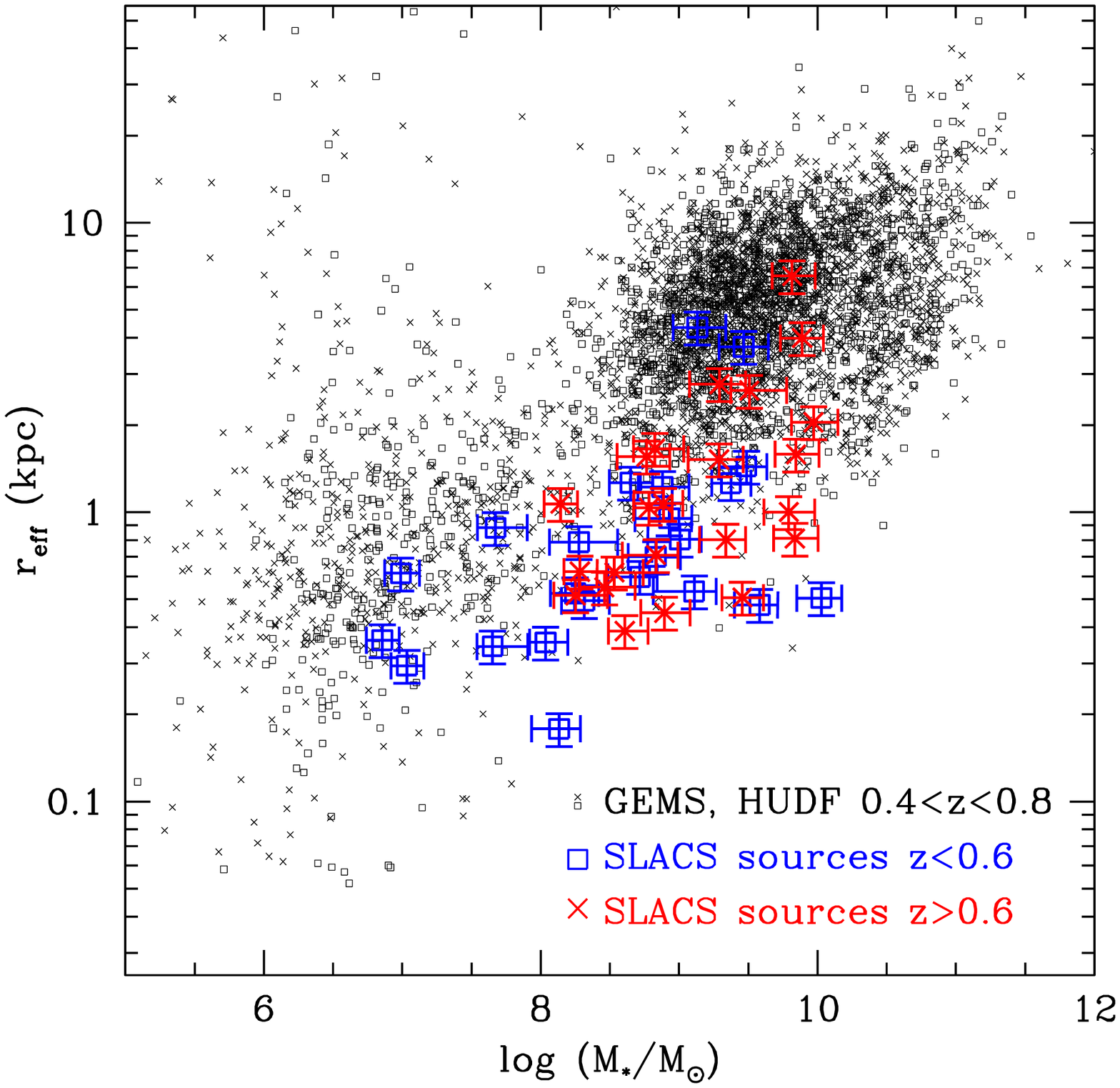}
\caption{Size--mass relation for SLACS source galaxies (large points with
error bars bars) and all GEMS and HUDF galaxies (small points).
Crosses represent galaxies between $0.4<z<0.6$ 
and squares those between $0.6<z<0.8$.  
Stellar masses for GEMS and HUDF galaxies have been computed using the
Bell et al. 2003 recipes. All stellar masses have been converted to a
Kroupa IMF normalization.
\label{fig:sizemass}}
\end{figure}

We note that, as expected
for a flux limited sample, the intrinsically fainter objects are only
detected at the lowest redshifts (in Figs.~\ref{fig:sizemag}-\ref{fig:sizemass}, 
crosses represent galaxies between $0.4 \lesssim z \lesssim 0.6$ 
and squares those between $0.6 \lesssim z \lesssim 0.8$). Other than this luminosity
segregation, no evolution is apparent within our dataset.

The color-magnitude diagram shown in Figure~\ref{fig:colormag} paints
a similar picture of the properties of the SLACS sources. They span
the range of blue colors typical of star-forming galaxies, and of the
GEMS and HUDF samples, even with their different intrinsic luminosities. 
Again, other than luminosity segregation, no evolution is apparent.
Our data (and GEMS) are also consistent with no relation between color and
magnitude, in contrast to the sample of HUDF galaxies we 
have selected for comparison, which do show bluer galaxies at 
higher redshift.  This perhaps surprising result is consistent with those of
\citet[]{Pir++06}, who looked at emission line galaxies (ELGs) in the HUDF,
with $-14 < M_B < -22$ and rest-frame B-V colors from $\sim0$ to
$\sim1$.  As we discuss below, the SLACS sources may be best
represented as ELGs, which could be the cause of this similarity.

Finally, the offset in size with respect to GEMS and HUDF is
pronounced even in the size-stellar mass plane
(Figure~\ref{fig:sizemass}), where the SLACS sources seem to define a
sequence offset to smaller radii by $0.5$~kpc, or $\sim30\%$.  This
indicates that the offset in the size-luminosity relation is not due
to abnormally low mass-to-light ratios but to an intrinsic compactness
of the population.  For the SLACS sources, we find a size-mass
relation of $\log_{10} \Reff \rm (kpc) = 0.24 \times
\log_{10}{M_{*}/\msun} - 2.20$, although we again caution the reader
as to the small sample size and uncertain selection function.

%
%

We note that the stellar masses of the SLACS sources are remarkably
low for a population at cosmological distances, extending below 10$^8$
M$_\odot$.  Three systems in particular have remarkably low stellar
mass:\newline
J1538$+$5817 ($\zs$=0.531, $M_{*}$=10$^{7.03}\msun$, $L_V$=10$^{7.9}\lsun$), 
J1402$+$6321 ($\zs$=0.481, $M_{*}$=10$^{6.99}\msun$, $L_V$=10$^{7.8}\lsun$),
J1719$+$2939 ($\zs$=0.578, $M_{*}$=10$^{6.86}\msun$, $L_V$=10$^{7.7}\lsun$).

Their $V$-band luminosities are just 0.5-0.8~dex higher than those of
the brighter Milky Way dwarf satellites, Sagittarius and Fornax, and
considerably fainter than the Large Magellanic Cloud
\citep[$3\cdot10^9\lsun$][]{vdM++02}.  As can be seen in Figure~\ref{fig:montage}, these
three sources are being lensed into high magnification partial
Einstein rings (their total magnifications are the three highest: 
44.3, 31.2 and 40.7 respectively). 
The J1538$+$5817 source may be a low luminosity
satellite of a brighter companion, visible just outside the ring.


\subsection{SLACS sources as emission line galaxies}
\label{sec:elgs}

Although the comparison between our galaxies and GEMS or HUDF is useful
-- it allows us to compare the physical properties of our galaxies,
including stellar mass, to large samples with the same redshift
distribution -- the selection algorithm for SLACS is quite
different from that of GEMS or HUDF.  
The SLACS sample was selected based on the detection of
source galaxy emission lines \citep[]{Bol++04};  GEMS and HUDF are both imaging
surveys.  It is therefore useful to compare the SLACS sample with
emission line selected galaxies (ELGs). SLACS is sensitive to observed
fluxes as low as $\sim$6$\cdot$10$^{-17}$erg s$^{-1}$cm$^{-2}$
\citep{Bol++04,Dob++08}. Taking into account the effects of
magnification the equivalent depth is typically
$\sim$7$\cdot$10$^{-18}$erg s$^{-1}$cm$^{-2}$, but can be higher by a
factor of a few for the most extreme objects.

Emission line galaxies identified by HST grism surveys
\citep[e.g.][]{Dro++05,Str++09} reach comparable depths, and provide a
good benchmark. Those samples span similar broad band magnitude ranges
to the SLACS sources and have comparable sizes. For example, the size
distribution of galaxies in the sample of \citet[]{Dro++05} is limited
to small sizes ($\Reff<0\farcs5$) and peaks at $\Reff \sim 0\farcs1-0\farcs2$.
Magnitudes peak at F814W$\sim$24, reaching magnitudes as
faint as 27.  These are quite similar distributions to those for the SLACS galaxies.

\section{Discussion and conclusions}
\label{sec:conc}

Following an initial foray by \citet{Mar++07}, we have extended the
study of the size-mass relation of galaxies to lower masses and
smaller sizes by making use of the factor of $\sim 10$ magnification
provided by the gravitational lenses of the SLACS survey. In order to
model this unprecedented sample of 46 gravitational lens systems in
three bands (\hstV, \hstI, \hstH), we have developed and tested
\klens, a new code for fast lens modeling optimized for this
work. Extensive testing and comparison with standard codes like
\galfit\  show that we are able to accurately measure the properties of
small, faint lensed galaxies: for a mock lensed galaxy of 25th
\hstI-band magnitude, we are able to recover a $0\farcs05$ size.

The main result of this study is that by exploiting the gravitational
lensing effect we have been able to determine sizes and stellar masses
of a sample of very faint and compact galaxies, which are typically
not studied by imaging surveys and are beyond the reach of normal
spectroscopic follow-up.  We can derive accurate sizes and stellar
masses for SLACS sources that are typically 1-2 magnitude fainter and
5 times smaller than those of a galaxies selected for structural
analysis by typical HST imaging survey (GEMS).  Although the lensed
sources are not as faint as galaxies from extremely deep imaging
surveys (HUDF), we are able to reach \hstI\ magnitudes of 27 with
limited exposure times.  For comparison, the HUDF survey is comprised
of 400 \hst\ orbits, with 144 in I-band, while the SLACS survey is
just one orbit in each band. Furthermore, we identify galaxies that
are typically $\sim30\%$ smaller in effective radius than those
identified in the HUDF at comparable luminosity or stellar mass.

The SLACS sources are similar to emission line
selected-galaxies \citep[e.g.][]{Koo++95,Phi++97,Dro++05,Str++09}, and
as such may represent the building blocks of present day
dwarf-spheroidals or perhaps even bulges of galaxies
\citep[see][]{Ham++01}. However, thanks to the lensing effect we can
image them at high angular resolution and derive accurate
sizes and structural parameters.

The existence of this population highlights the importance of diverse
selection criteria for a complete census of the galaxy population. 
Lensing imposes a completely different selection function than for
imaging surveys: in particular, the SLACS sources were selected by
their strong, multiple emission lines, which were observed to stand out
against the background of an early-type (lens) galaxy spectrum
\citep{Bol++04}.  In contrast, the GEMS and HUDF sources were selected on the
basis of their broad band flux and morphology, using no information as
to their brightness in emission lines. Similarly, it is likely that
spectroscopic lens surveys could preferentially select compact
galaxies, owing to the larger magnification for compact
sources. Modeling the selection function in detail \citep[for example
along the lines suggested by][]{Dob++08} is essential to quantify the
luminosity function of this population and its abundance relative to
the population of broad-band selected galaxies, and that of emission
line selected galaxies. This modeling work goes beyond the scope of
this paper but will be pursued in the future with the goal of
determining the shape an intrinsic scatter of the size-mass relation
taking into account the lensing-selected population.

A brief summary of the main results of this paper follows:

\begin{itemize}

\item{Our best-fit lens mass parameters are in good agreement with those 
found in \paperV\ and \paperIX\ in the \hstI\ band. To mitigate
against the scatter introduced by lens light subtraction in the other,
lower S/N filters, we fixed the mass models to those fitted in the
highest S/N band.}

\item{The inferred sources appear to be similar in structure to the galaxies
in the GEMS catalog of \citet{Bar++05} and the HUDF catalog of 
\citet[]{Coe++06}, in the sense that the
distributions of \sersic\  indices are the same between the three
samples. However, the SLACS sources have a significantly different 
distribution of magnitudes.  They are fainter than GEMS, with the peaks 
of their differential number counts separated by 2 magnitudes, and 
brighter than HUDF by about 4 magnitudes.  (However, SLACS reaches 
these magnitudes with only one \hst\ orbit in I-band, while HUDF uses 144).  
Aside from this shift, the color-magnitude diagrams of
the three samples are consistent with each other.}

\item{The SLACS sources are also significantly smaller than the GEMS galaxies
and have sizes similar to the HUDF galaxies.
The SLACS sources are made measurable via the lens magnification, while the
extremely long exposure time is what allows study of the HUDF galaxies. The size
distributions of the SLACS sources and the HUDF galaxies peak at approximately 0.1~arcsec ($\simeq 0.8$\,kpc); such
small, faint objects were likely below the COMBO-17 flux limit used in
constructing the GEMS sample, or were unresolved in the shallow \hst\ images.}

\item{The closest analog in size and magnitude to the SLACS sources
are the faint emission line galaxies identified by blind grism surveys
with \hst.}

\item{Combining the reconstructed (model) magnitudes from the three
\hst\ bands, we infer stellar masses for each SLACS source and plot the
size-mass relation at $0.4 < z < 0.8$.  We then compare to the GEMS and HUDF
results, finding that the SLACS sources are offset from the GEMS and HUDF
size-mass and size-luminosity relations towards smaller effective radii.}

\item{Some of our measured objects are very low stellar  mass indeed, $\sim
10^7\msun$, comparable to the largest dwarf satellites of the Milky
Way and fainter than the Large Magellanic Cloud. }

\end{itemize}

This work is a further step towards extending the study of
intermediate and high redshift galaxies to smaller and fainter
galaxies, demonstrating that gravitational lensing can help us probe
this regime.  A larger sample of lenses will improve the statistical
significance; the analysis code we have developed can perform lens
modeling simply, quickly, and robustly and will be useful when even
large samples of lenses are available. Before this, a fuller
understanding of the lensed source selection function will allow us to
make a quantitative analysis of the size-mass relation of dwarf
galaxies at intermediate redshift and how this relation evolves with cosmic
time.


\acknowledgments

We wish to thank Scott~Burles for all his contributions as one of
the ``founding members'' of SLACS collaboration.  We also thank
Laura~D.~Melling for her initial efforts in developing \klens.
TT and ERN acknowledge support from the NSF through CAREER award
NSF-0642621, and from the Packard Foundation through a Packard
Research Fellowship.  ERN is currently supported by a National 
Science Foundation Graduate Research Fellowship.
PJM is grateful to the TABASGO and Kavli foundations for support in
the form of two research fellowships.
RG acknowledges support from the Centre National des \'Etudes Spatiales.
LVEK is supported in part through an NWO-VIDI
program subsidy (project number 639.042.505).
The work of LAM was carried out at the Jet Propulsion Laboratory, 
under a contract with NASA.  LAM acknowledges support from the 
NASA ATFP program.
This research is supported by NASA through Hubble Space Telescope
programs SNAP-10174, GO-10494, SNAP-10587, GO-10798, GO-10886, and in
part by the National Science Foundation under Grant No. PHY99-07949,
and is based on observations made with the NASA/ESA Hubble Space
Telescope and obtained at the Space Telescope Science Institute, which
is operated by the Association of Universities for Research in
Astronomy, Inc., under NASA contract NAS 5-26555.


\appendix

\section{Measuring Source Galaxies with \klens\ }
\label{sec:klens}


We aim to measure the magnitude and effective radius of the source
galaxies of the gravitational lenses in our sample. In this section we
describe a new code, \klens, designed for fast and robust lens
modeling of large number of images (for this paper alone we need to
model 138 independent images, i.e. 3 bands for each of the 46
lenses). It is similar in spirit to the software used by
\citet{Mar++07}, in that the models for the lens mass and source
surface brightness distributions are the same. However, we do not
explore the posterior probability distribution (PDF) for the model
parameters fully, but instead simply search for the peak of the
distribution. This approach saves a considerable amount of CPU time
(up to two orders of magnitude), since the computation of the image
pixel data likelihood function is relatively expensive -- but it gives
us only the covariance matrix in the Laplace approximation as opposed
to full statistical uncertainties on the parameters. We note however,
that the error budget in this kind of analysis is totally dominated by
systematic errors due to factors such as lens light subtraction, model
simplicity and PSF approximation \citep[see the following sections,
and][]{Mar++07}. Therefore we conclude that, with current CPU
limitations, it is more effective to combine minimization with Laplace
approximation of the errors with a detailed study of one or more
objects to estimate systematic errors \citep{Mar++07}.

After describing the lens and source models in a bit more detail, we
outline our implementation of the prior PDFs and the optimization of
the posterior, before demonstrating the code's performance on three
test datasets.


\subsection{Models}
\label{sec:klens:model}

The gravitational lenses in our sample are known to be
well-approximated by simple, singular isothermal ellipsoid (SIE) mass
distributions \citep[]{Koo++06}.  We therefore use an SIE model to
describe the mass distribution of the lens galaxy.  This model has
five parameters: the velocity dispersion $\sigma_{\rm SIE}$,
elliptical axis ratio $q = b/a$, orientation angle $\theta_{\rm d}$,
and position ($x_{\rm d},y_{\rm d}$). \citep[See \eg][for
details]{KSB94}.  For the source galaxy surface brightness, we use an
elliptically symmetric distribution with \sersic\  profile. The source
model has seven parameters: the position ($x_{\rm s},y_{\rm s}$),
orientation angle $\theta_{\rm s}$, inclination angle $i$ (more often
referred to be $\cos{i}$, which equals the apparent ellipse axis
ratio), effective radius $\Reff$, AB apparent magnitude $m$, and
\sersic\  index $n$.  A \sersic\  index of $n=0.5$ corresponds to a
Gaussian, $n=1$ to an exponential disk, and $n=4$ to a de Vaucouleurs
profile \citep[see \eg][for details]{Pen++02}.

In \paperV\ we used multiple \sersic\  profile components to describe the
source galaxy.  Because our goal here is to measure the physical
properties of the source such as size, magnitude and mass, and compare
them with corresponding measurements of non-lensed galaxies in the
literature, we choose instead to describe the source as a single
object.  As a result, we do not expect to be able to fit all our
sources to the same level of detail as we did in \paperV.  One option
would be to remove all lenses with complex sources from our sample;
however this would bias us towards smaller and simpler sources, which
is undesirable. This same single-component simplification was used in,
for example, the GEMS and SDSS size-mass relation studies; restricting
ourselves to single-component sources allows us to to make direct
comparisons with these nonlensing surveys.

Asserting that the assumption of an SIE model for the lens galaxy mass
profile is the most significant source of systematic uncertainty,
\citet{Mar++07} estimated the systematic errors in the source size and
brightness to be approximately 12\% ($0.07$~kpc for a $0.6$~kpc source
at $\zs$=0.6) and $0.26$~magnitudes respectively. However, we note
that these errors will be common to all observations of a given lens.


\input{tab3.tex}

\subsection{Priors}
\label{sec:klens:priors}

To ease the exploration of the model parameter space, we work with
hidden parameters constrained to vary uniformly on a hypercube,
such that their typical values are around one. These hidden parameters
are then transformed to the physical parameters described in the
previous section before the predicted image is generated. This
transformation need not be linear: in fact, it is used to encode our
prior knowledge about the model parameters. For example, a uniform
prior PDF is implemented as a linear transformation, while
exponentiating a hidden parameter corresponds to assigning a
scale-free (``Jeffreys'', $1/x$) prior.  Truncation of the prior PDFs
is achieved where necessary by setting the log likelihood to be very
large and negative outside the specified limits.  The parameters of
our model are listed in Table
\ref{tab:models}, along with the prior PDFs assigned to them, unit
transformation used, and the range over which they are allowed to vary.


\subsection{Posterior evaluation and optimization}

At each point in parameter space visited we compute the pixel value
likelihood function. We assume Gaussian errors on the pixel values,
given by the weight (inverse variance) map produced during data
reduction (\paperIX). The log likelihood is then just minus half the
image $\chi^2$. The predicted image, required for this statistic,
is computed by mapping each pixel's position back to the source plane
via the lens equation, and looking up the value of the model source
surface brightness. We over-sample the image plane to make sure we
resolve the source. The lensed images are then convolved with a model
PSF; we use Tiny Tim \citep{TinyTim} to generate this PSF.  Note that we 
do not model lens galaxy light, only the source surface brightness.
We therefore assume perfection subtraction of lens galaxy light, an assumption
which incurs the systematic error discussed at the end of \secref{sec:slacsintro}


Since the prior PDF is implemented by parameter transformation,
optimization of the posterior is equivalent to maximizing the log
likelihood with respect to the hidden parameters.  For this
optimization we use the IDL routine
\mpfit\ \citep{MPFIT}, which is a generalized implementation of the
Levenberg-Marquardt nonlinear least-squares algorithm.

The likelihood function is typically quite sharply peaked, with
multiple local maxima to be avoided. We find that a three-step process
is required for the global peak to be successfully located.  First, we
make an initial estimate of the source galaxy parameters using a
best-guess mass model: the image pixels are traced through the mass
model using the lens equation and flux is accumulated on a grid in the
source plane. Where multiple image pixels map to the same source plane
pixel, we take their simple average. We then take the initial source
galaxy position to be the first moment of this resulting light
distribution, while summing all the source plane pixel values provides
an initial approximation to the source magnitude (this is usually an
over-estimate). We gain no knowledge of the source inclination, size,
position angle or \sersic\  index in this first stage; the source
position and magnitude are the most important for finding a good fit
to the data, in the sense that small deviations in these parameters
can lead to very low likelihood models.

Next, we refine our initial estimates by running \mpfit, using our
initial estimates of the source position and magnitude as its starting
point (we initialize the other parameters at typical values).  This
preliminary run is performed without image plane subsampling, for
speed.  We also use a deflated weight image (the square root of the
weight) in this second stage; this smooths the likelihood function,
making it easier for the optimizer to locate the global maximum.
Finally, in the third step we re-start \mpfit\ at the position found in
step two to get final maximum posterior model parameters for the mass
and source. For this final run, we use an image plane over-sampling of
4 (along each axis), and the normal weight image.  Convolution with the PSF is
done on the over-sampled grid; we then use neighborhood averaging to
return to the original size.


\subsection{Testing \klens\ }
\label{sec:klens:testing}

In order to verify that the \klens\  code works properly, we performed
tests on (1) simulated non-lensed galaxies, (2) real non-lensed
galaxies and (3) simulated lenses.  All simulated data was created
using \hst/ACS \hstI-band (F814W) image parameters and PSFs, while the
real galaxies were selected from
\hst/ACS  \hstI-band images.

\subsubsection{Testing \klens\  on simulated non-lensed galaxies}
\label{sec:klens:testing:nonsim}

\input{tab4.tex}

\begin{figure*}
\centering
\begin{minipage}{0.48\linewidth}
\includegraphics[width=0.9\linewidth]{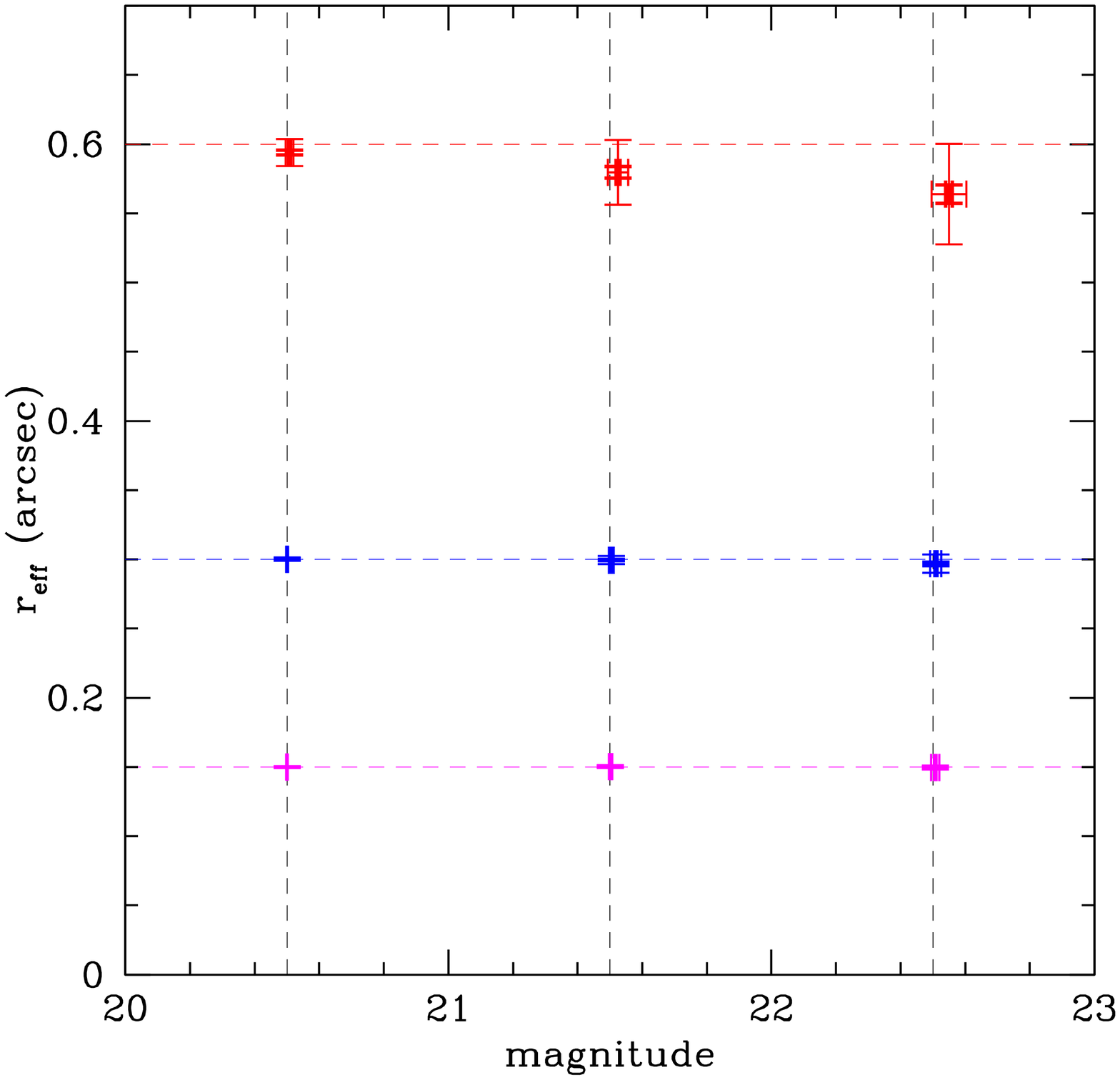}
\end{minipage}\hfill
\begin{minipage}{0.48\linewidth}
\includegraphics[width=0.9\linewidth]{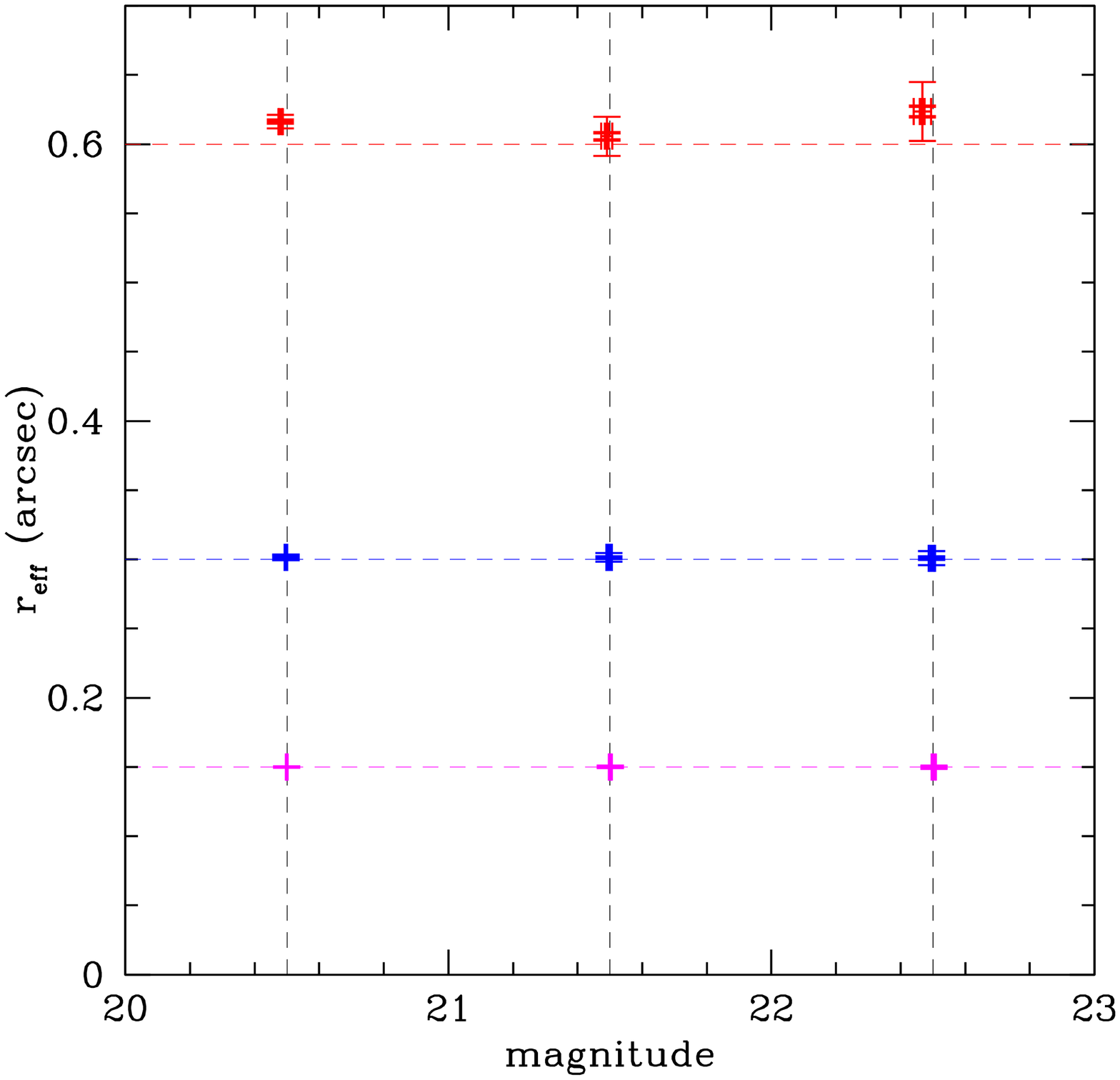}
\end{minipage}
\caption{Testing \klens\  and \galfit\  on simulated non-lensed galaxies.  Best-fit
effective radii and magnitudes for \klens\  (left panel) and \galfit\  (right
panel), binned by true size and magnitude: the dashed vertical and horizontal
lines indicate these bins.  Errors bars indicate the r.m.s. scatter in each bin
(the errors on the mean are $\sim5.5$ times smaller and are not shown).  In the
largest size bins, \klens\  tends to underestimate size and overestimate
magnitude, while \galfit\  tends to do the opposite. \label{fig:nonsim}}
\end{figure*}

We first test \klens\  on simulated non-lensed galaxies. Each mock
galaxy is taken to be a single elliptically symmetric \sersic\  profile
component, to allow us to investigate straightforward parameter
recovery by the code. The sizes, magnitudes and \sersic\  indices of
these galaxies were varied over a 3 by 3 by 3 grid: we chose sizes
from $(0\farcs15, 0\farcs3, 0\farcs6)$, magnitudes from $(20.5, 21.5, 22.5)$
and \sersic\  indices from $(0.5, 1, 2)$. All position angles were set to
$\thetas=1.75$ and inclinations at $\cos(i)=q=0.6$.  The simulated
galaxies were placed at the center of $42\times42$~pixel images.
Magnified by a typical lens, this size is approximately equivalent to
the $122\times122$~pixel cutouts used for the lenses. We generated ten
independent noise realizations for each galaxy, giving a total of 270
simulated observations.  The whole sample were then modeled with both
\klens\  (using the priors given in Table~\ref{tab:models}) and \galfit.

\begin{figure}
\centering
\includegraphics[width=0.9\linewidth]{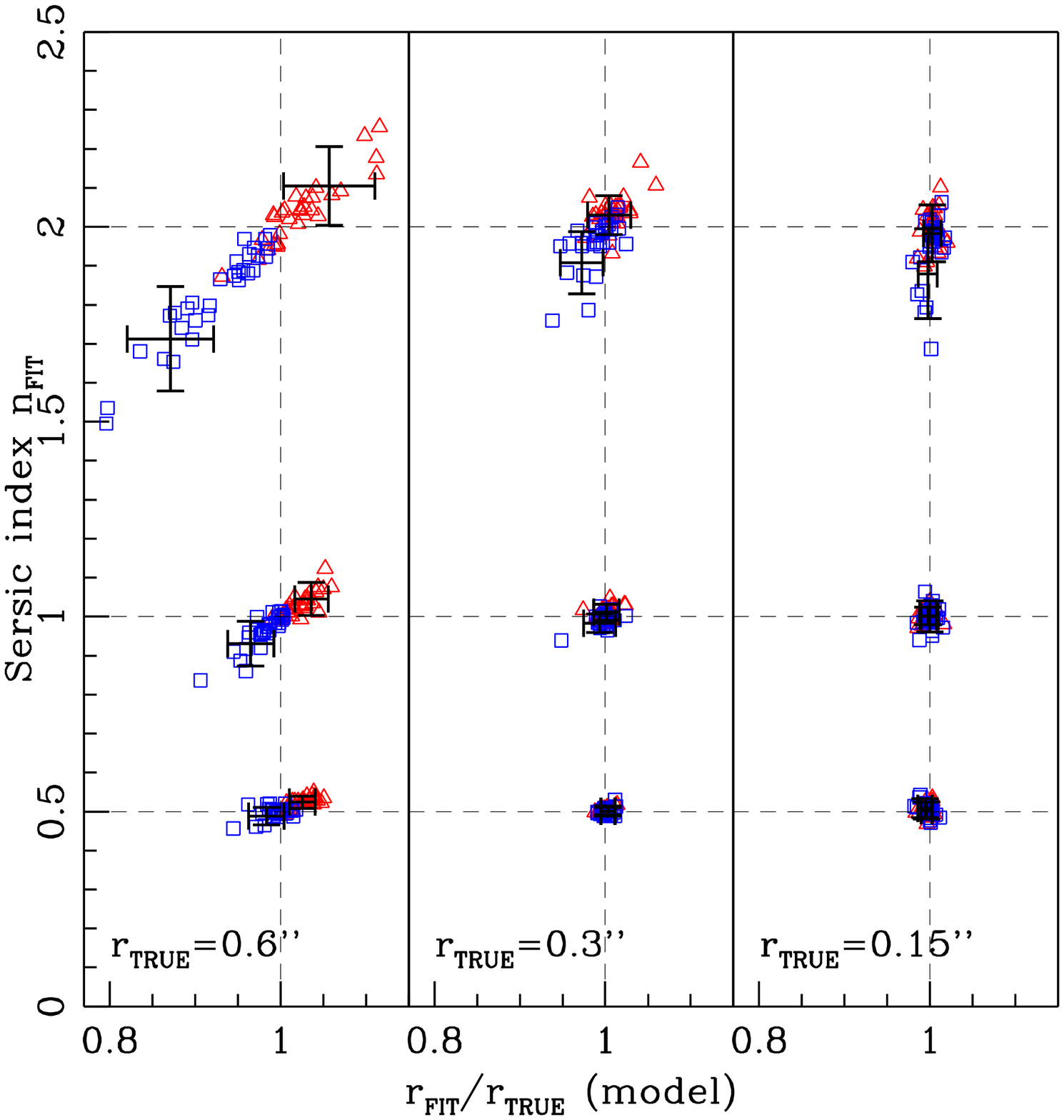}
\caption{Testing \klens\  and \galfit\  on simulated non-lensed galaxies.  We plot
best-fit \sersic\  index against the ratio of the best-fit to true size, in order to
illustrate the covariance between \sersic\  index, size and
magnitude. Galaxies are again binned by true \sersic\  index and size,
shown by the dashed lines.  Red triangles indicate \galfit\  fits and
blue squares indicate \klens\  fits.  The error bars indicate the mean
value and scatter of galaxies in the \emph{faintest} bin (the errors
on the mean are $\sim3$ times smaller and are not shown).  The
covariance -- and the fact that \klens\  and \galfit\  sample disparate
parts of it -- is most evident in the $n=2$, $r=0\farcs6$
bin. \label{fig:degen}}
\end{figure}

The results of the parameter recovery are given in
Table~\ref{tab:nonsim}, which lists the mean (inferred -- truth)
difference, and associated r.m.s. scatter values.  We look at the
results more closely in Figure~\ref{fig:nonsim}.  Overall \klens\  and
\galfit\  give very similar results, although the rms scatter is larger with
\klens. For both modeling programs, the
largest galaxies contribute most of the bias, which is in the sense
that \klens\  tends to slightly underestimate sizes, while \galfit\  tends
to slightly overestimate them, even though the residual images are
consistent with simulated noise.  This is because the two programs are
sampling different parts of the covariance between size, magnitude and
\sersic\  index.  To illustrate this point, we plot the best-fit \sersic\ 
indices against the best-fit sizes in Figure~\ref{fig:degen}, again
grouped by true parameter value. 

\subsubsection{Test on real non-lensed galaxies}
\label{sec:klens:testing:nonreal}

Next, we tested \klens\  on real non-lensed galaxies.  We selected a
sample of 981 galaxies, detected in a subset of the SLACS fields, to
match the nine magnitude and size bins used in the previous test: the
sample consists of all objects in those fields that have \sextractor
magnitudes $20<m<23$ and $0\farcs1<r<0\farcs65$ that are not stars or image
defects.  We then ran \klens\  on each test image, with the priors given
in Table~\ref{tab:models}; we also ran \galfit\  on each dataset.

\begin{figure}
\centering\includegraphics[width=0.9\linewidth]{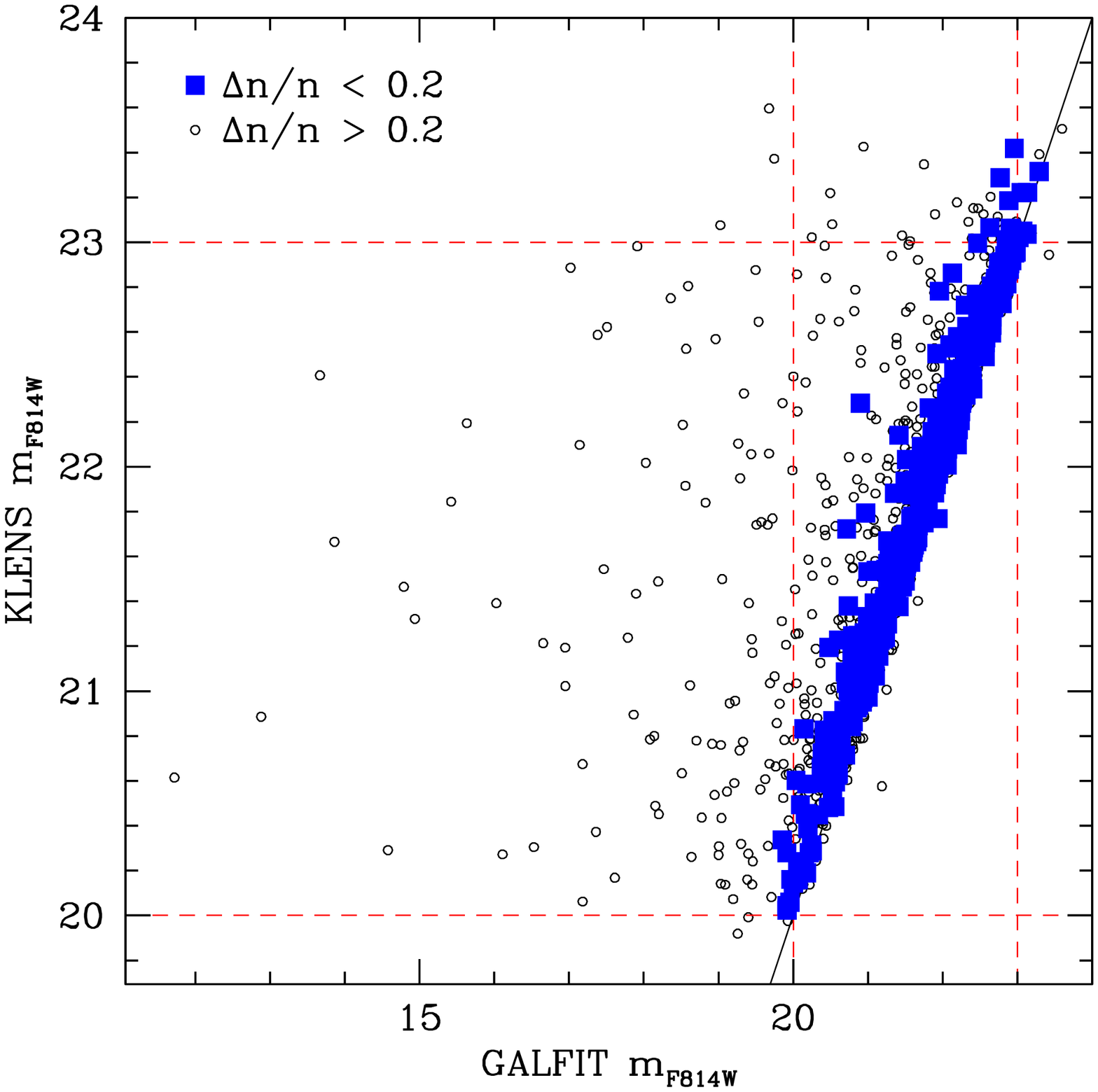}
\caption{Testing \klens\  against \galfit\  for real galaxies. 
Shown are the best-fit magnitudes from \galfit , on the $x$-axis, and
from \klens, on the $y$-axis.  Large filled blue squares are those
points for which the percent difference in \sersic\  index is less than
20\%. The red dashed lines indicate the upper and lower limits on the
\sextractor\ magnitudes for objects in the
sample. \label{fig:nonreal_m}}
\end{figure}

For this sample we do not know the true values for each object's size,
magnitude and \sersic\  index; we can only compare \galfit\  and \klens\ 
results.  When comparing \sersic\  indices, we noted significant
systematic differences between the two programs, especially at
$n>1.5$.  This scatter is, as discussed in
\secref{sec:klens:testing:nonsim}, a result of \klens\  and \galfit\ 
sampling different parts of the covariance between \sersic\  index, size
and magnitude.  We compare the two codes' best-fit magnitudes in
Figure~\ref{fig:nonreal_m}.  In this plot we highlight those galaxies
for which best-fit \sersic\  indices differ by less than 20\%; 408
objects meet this requirement. The magnitudes and sizes for these
galaxies were found to to be in reasonable agreement: $\langle\Delta
m\rangle = 0.147\pm0.008$ with r.m.s. scatter $0.170$.  A similar
analysis for size gives $\langle \Delta \Reff\rangle =-0.174\pm0.009$
with r.m.s. scatter $0.175$.  As we decrease the allowable difference
in \sersic\  index, the bias and scatter both decrease.

\galfit\  occasionally returns very low magnitudes relative to the \sextractor\
values, and that these are associated with \sersic\  indices very
different from the \klens\  values. We note that, for example,
\citet{Bar++05} reject objects with \galfit\  magnitudes differing from their 
\sextractor\ magnitudes by more than 0.6 magnitudes. 

\subsubsection{Test on a simulated lens sample}
\label{sec:klens:testing:synthetic}

Finally, we tested the ability of \klens\  to recover the structural
parameters of lensed source galaxies.  We can do this to a limited
extent by comparing our \klens\  mass models to the original SLACS lens
models (\secref{sec:slacs:cfpaperV}). However, for the source
parameters we need to use simulated galaxies, as we did in
\secref{sec:klens:testing:nonsim} in the non-lensed case.  We
generated a sample of lensed galaxies to be representative of the
possible lensed image configurations and sources. The exact properties
of the lens itself are less important: we used the same mass profile
($\sigmaSIE$=270~km s$^{-1}$, $q$=0.77 and $\thetad$=1.75~rad, typical
parameters for a SLACS lens) for each synthetic system.  We also fixed
the source galaxy orientation at $\thetas$=1.75~rad and its
inclination at $\cos(i)=q=0.6$, again typical values.

The source position, effective radius, magnitude and \sersic\  index were
varied, to give 108 possible lenses.  We chose four source positions
such that the four main lens morphologies (double, caustic, cusp and
quad) were represented. The source effective radius was chosen from
$(0\farcs05, 0\farcs1, 0\farcs2)$, the magnitude from $(23, 24, 25)$ and the
\sersic\  index from $(0.5, 1.0, 2.0)$.  Taking into account
magnification due to lensing, these sizes and magnitudes are roughly
equivalent to those used in the non-lensed galaxy tests.  For each
mock system we simulated pixelated images, again with no lens galaxy
light. We generated ten noise realizations for
each lens, to give a total of 1080 mock observations. We then ran
\klens\  on each mock dataset, with the priors given in
Table~\ref{tab:models}.

\begin{figure}
\centering\includegraphics[width=0.9\linewidth]{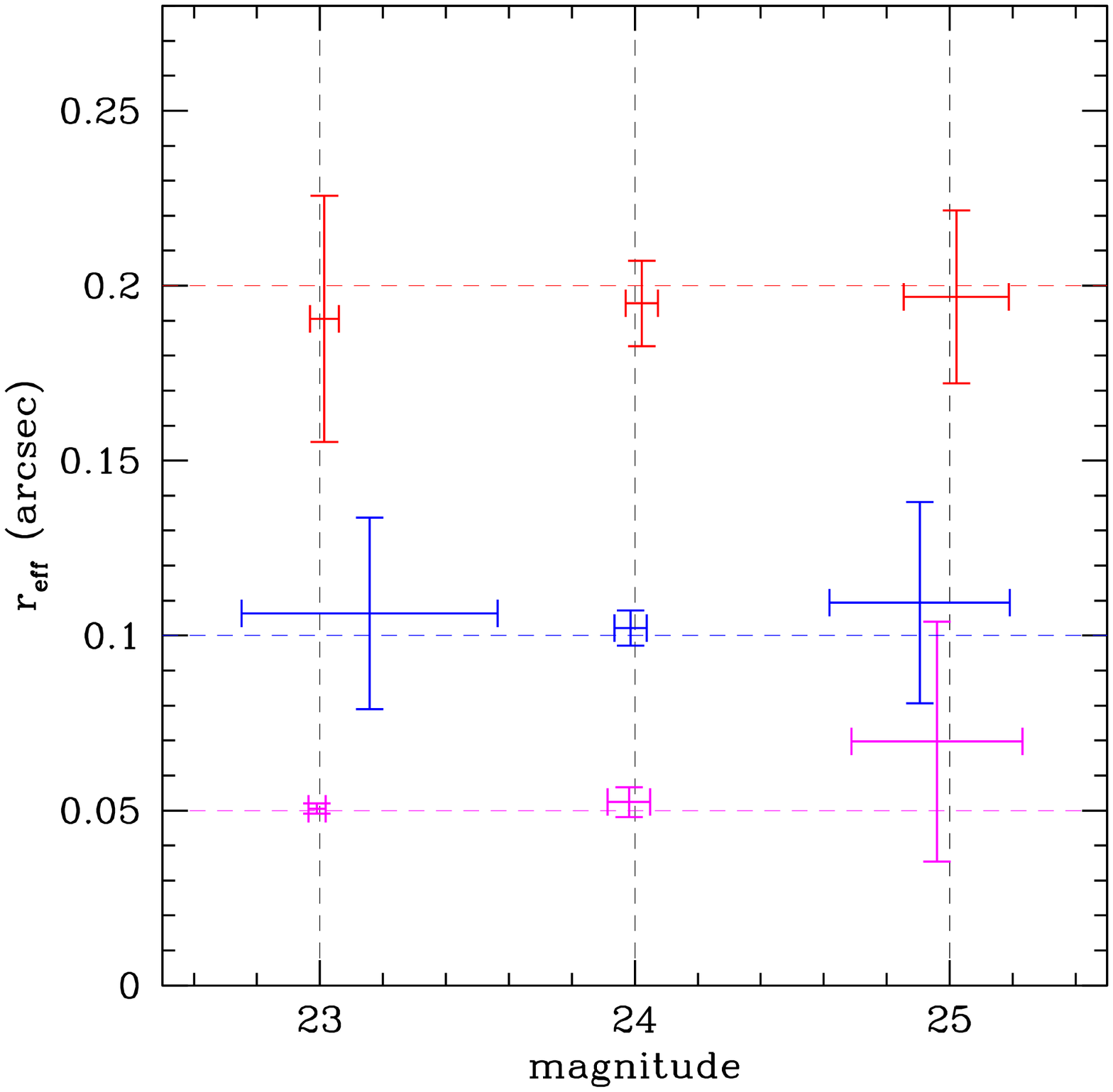}
\caption{Testing \klens\  source parameter recovery on simulated lensed galaxies. 
We plot \klens-inferred model magnitude ($x$-axis) and effective radii
($y$-axis) for a sample of 1080 simulated lenses.  The faint dashed
lines again indicate the true parameter values (the 9 possible
combinations of true size and magnitude occur at their intersections).
The data is binned by true size and magnitude; the nine points show
the average values for each bin.  The error bars again indicate the
r.m.s. scatter of the inferred parameters (the errors on the mean are
11 times smaller and are not shown). \label{fig:ReMag2}}
\end{figure}

Parameters for which the same true value was used in all simulations
(i.e. the lens parameters, and the source inclination and orientation
angle) are well-modeled, with only small biases well below the
r.m.s. scatter (see Table~\ref{tab:synfix}). We divide the sample into
bins to consider the inferences of the remaining parameters, their
means and scatters across the sample; these results are presented in
Table~\ref{tab:synthetic}.  In Figure~\ref{fig:ReMag2} we look at the
performance of \klens\  in recovering the input effective radius and
magnitude, binned by their true values.

We find that the inferred parameters are consistent with the input
values, albeit with larger scatter than was seen in the non-lensed
simulation case. This is partly due to our use of fainter galaxies,
which are harder to measure when the image configuration is of lower
magnification. (We note that the scatter is smaller for high surface
brightness galaxies, indicating that -- as might be expected -- it is easier to model such
galaxies.) It may also be due to the way the information in the data
is being used to simultaneously constrain the lens model -- to some
extent changes in the source parameters can be balanced by changes in
the lens model.

\input{tab5.tex}
\input{tab6.tex}

\subsubsection{Summary}
\label{sec:klens:testing:summary}

We have tested \klens\  on three different types of data. Our results on
simulated, non-lensed simple \sersic\ profile galaxies show that \klens\ 
gives almost identical results to \galfit; there is a small ($\lesssim
5$\% for 0\farcs6 size objects) systematic bias towards larger inferred
sizes and \sersic\  indices with \galfit, and a bias of comparable
magnitude towards smaller inferred sizes and \sersic\  indices with
\klens, as they sample different parts of the covariance. In
real, non-lensed galaxies, this covariance results in significant,
systematic differences between \klens\  and \galfit\  inferred parameters,
with \galfit\  suffering more catastrophic magnitude errors -- these
have simply been removed in the literature. The difference between the
codes is less pronounced if only models with matching \sersic\  indices
are considered. Finally, testing \klens\  on simulated lens systems
where the source is a simple \sersic\ profile galaxy shows that, for
typical expected sizes and magnitudes, \klens\  is able to provide
accurate measurements of these quantities, with its precision
improving with source surface brightness. From this last test
(Figure~\ref{fig:ReMag2}), we estimate that the statistical
uncertainties on the source $\Reff$ and brightness are 5\% and
0.1~magnitudes respectively. Adding these to the other systematic
errors noted above in quadrature, we find overall systematic errors of
13\% on $\Reff$, and 0.3~magnitudes photometric error (although
0.26~mag in quadrature of the latter -- due to assuming an SIE for lens 
mass model -- is common to all filters studied).


\bibliographystyle{apj}


\end{document}

%% file: tab1.tex
\begin{table*}
\begin{center}
\caption{\label{tab:obs}}
\begin{tabular}{lccccccccc}
\hline\hline
Name  &  $\zd$  &  $\zs$  & \hstV  &  \hstI &  \hstH  &   Instrument  & $n$  & $\Reff$ & $\mu$ \\ 
      &         &         &               &           &          &              &       &   (arcsec)  &   \\ \hline 
J0037$-$0942  &  0.195  &  0.632  &  23.96  &  23.59  &  23.70  &  ACS*  &  1.46  &  0.06  &  5.8  \\
J0044$+$0113  &  0.120  &  0.196  &  21.44  &  21.05  &  20.79  &  ACS*  &  0.90  &  0.16  &  3.6  \\
J0157$-$0056  &  0.513  &  0.924  &  26.61  &  25.45  &  23.16  &  ACS  &  1.70  &  0.83  &  2.7  \\
J0216$-$0813  &  0.332  &  0.524  &  24.46  &  23.25  &  22.77  &  ACS  &  0.33  &  0.69  &  3.1  \\
J0330$-$0020  &  0.351  &  1.071  &  23.46  &  22.73  &  22.30  &  ACS  &  1.33  &  0.12  &  4.0  \\
J0737$+$3216  &  0.322  &  0.581  &  25.16  &  24.04  &  23.62  &  ACS  &  1.08  &  0.11  &  15.5  \\
J0912$+$0029  &  0.164  &  0.324  &  25.70  &  24.92  &  23.34  &  ACS  &  0.20  &  0.27  &  10.3  \\
J0935$-$0003  &  0.347  &  0.467  &  23.91  &  23.45  &  22.97  &  ACS*  &  0.81  &  0.16  &  2.4  \\
J0955$+$0101  &  0.111  &  0.316  &  22.32  &  22.44  &  20.96  &  ACS*  &  0.90  &  0.31  &  3.1  \\
J0959$+$0410  &  0.126  &  0.535  &  24.46  &  22.68  &  21.36  &  ACS  &  2.49  &  0.08  &  6.0  \\
J1100$+$5329  &  0.317  &  0.858  &  25.55  &  25.42  &  25.74  &  ACS*  &  1.14  &  0.14  &  20.6  \\
J1103$+$5322  &  0.158  &  0.735  &  25.33  &  24.24  &  22.45  &  ACS  &  1.79  &  0.11  &  7.6  \\
J1106$+$5228  &  0.095  &  0.407  &  25.01  &  24.60  &  23.66  &  ACS*  &  0.20  &  0.11  &  28.0  \\
J1112$+$0826  &  0.273  &  0.630  &  23.32  &  23.25  &  23.14  &  ACS*  &  1.54  &  0.15  &  3.7  \\
J1142$+$1001  &  0.222  &  0.504  &  24.74  &  24.49  &  24.27  &  ACS  &  0.82  &  0.09  &  5.2  \\
J1143$-$0144  &  0.106  &  0.402  &  23.78  &  23.46  &  22.43  &  ACS  &  0.28  &  0.15  &  10.4  \\
J1204$+$0358  &  0.164  &  0.631  &  24.08  &  23.70  &  23.30  &  ACS*  &  2.06  &  0.24  &  7.9  \\
J1205$+$4910  &  0.215  &  0.481  &  24.91  &  24.14  &  24.63  &  ACS  &  1.76  &  0.06  &  13.9  \\
J1213$+$6708  &  0.123  &  0.640  &  25.71  &  25.42  &  24.82  &  ACS  &  1.99  &  0.07  &  10.1  \\
J1218$+$0830  &  0.135  &  0.717  &  25.08  &  24.30  &  23.25  &  ACS  &  1.17  &  0.11  &  4.2  \\
J1250$-$0135  &  0.087  &  0.353  &  24.36  &  23.61  &  21.90  &  ACS*  &  6.15  &  0.25  &  13.3  \\
J1250$+$0523  &  0.232  &  0.795  &  27.01  &  26.40  &  23.42  &  ACS  &  0.33  &  0.07  &  27.9  \\
J1306$+$0600  &  0.173  &  0.472  &  24.43  &  24.03  &  23.96  &  WFPC2  &  0.29  &  0.08  &  6.5  \\
J1313$+$4615  &  0.185  &  0.514  &  23.43  &  22.93  &  22.04  &  WFPC2  &  0.20  &  0.60  &  3.2  \\
J1318$-$0313  &  0.240  &  1.300  &  24.80  &  23.94  &  23.46  &  WFPC2  &  3.00  &  0.31  &  8.4  \\
J1319$+$1504  &  0.154  &  0.606  &  23.38  &  22.33  &  21.41  &  WFPC2  &  1.22  &  0.30  &  2.6  \\
J1402$+$6321  &  0.205  &  0.481  &  27.34  &  26.77  &  27.30  &  ACS  &  3.94  &  0.10  &  31.2  \\
J1416$+$5136  &  0.299  &  0.811  &  24.05  &  24.21  &  23.01  &  ACS  &  0.94  &  0.37  &  5.7  \\
J1420$+$6019  &  0.063  &  0.535  &  24.68  &  23.55  &  22.13  &  ACS  &  2.04  &  0.08  &  15.9  \\
J1432$+$6317  &  0.123  &  0.664  &  23.96  &  23.40  &  23.26  &  ACS  &  0.83  &  0.06  &  5.7  \\
J1436$-$0000  &  0.285  &  0.805  &  24.49  &  24.11  &  23.21  &  ACS*  &  3.17  &  0.20  &  4.8  \\
J1443$+$0304  &  0.134  &  0.419  &  25.24  &  25.33  &  24.99  &  ACS*  &  1.04  &  0.06  &  8.8  \\
J1451$-$0239  &  0.125  &  0.520  &  25.68  &  25.15  &  24.62  &  ACS  &  1.23  &  0.03  &  9.1  \\
J1525$+$3327  &  0.358  &  0.717  &  25.61  &  25.57  &  24.16  &  ACS  &  0.55  &  0.15  &  5.5  \\
J1531$-$0105  &  0.160  &  0.744  &  25.41  &  25.26  &  24.72  &  ACS*  &  1.14  &  0.09  &  12.4  \\
J1538$+$5817  &  0.143  &  0.531  &  25.92  &  26.32  &  28.96  &  ACS*  &  0.49  &  0.05  &  44.3  \\
J1621$+$3931  &  0.245  &  0.602  &  25.15  &  24.97  &  24.27  &  ACS  &  2.10  &  0.08  &  8.8  \\
J1630$+$4520  &  0.248  &  0.793  &  26.29  &  25.44  &  24.67  &  ACS  &  1.38  &  0.21  &  11.0  \\
J1644$+$2625  &  0.137  &  0.610  &  24.89  &  24.66  &  23.72  &  WFPC2  &  2.44  &  0.11  &  13.4  \\
J1719$+$2939  &  0.181  &  0.578  &  27.70  &  27.00  &  28.70  &  WFPC2  &  0.20  &  0.05  &  40.7  \\
J2238$-$0754  &  0.137  &  0.713  &  25.00  &  24.17  &  24.13  &  ACS  &  0.94  &  0.09  &  15.1  \\
J2300$+$0022  &  0.228  &  0.463  &  26.56  &  25.83  &  25.51  &  ACS  &  0.93  &  0.15  &  14.0  \\
J2302$-$0840  &  0.090  &  0.222  &  23.00  &  24.12  &  22.27  &  ACS*  &  0.28  &  0.22  &  13.0  \\
J2303$+$1422  &  0.155  &  0.517  &  25.13  &  23.81  &  23.28  &  ACS  &  0.21  &  0.20  &  8.0  \\
J2341$+$0000  &  0.186  &  0.807  &  24.04  &  23.49  &  22.13  &  ACS  &  0.86  &  0.21  &  10.9  \\
J2347$-$0005  &  0.417  &  0.715  &  24.09  &  23.96  &  21.81  &  ACS*  &  3.47  &  0.55  &  3.2  \\
\hline
\end{tabular}
\end{center}
{\footnotesize Notes:
AB magnitudes are from the Sersic model fits in the source plane, and so are 
unlensed. The effective radius $\Reff$ is the major axis 
of the elliptical isophote containing half the total flux. 
Uncertainties on $\Reff$ and the photometry are
13\% and 0.3 magnitudes, respectively.  For full coordinates, see \paperIX.
A star in the instrument column indicates that only a snapshot was available in \hstI (F814W).}
\end{table*}

%% file: tab2.tex
\begin{table}
\begin{center}
\caption{\label{tab:inf}}
\begin{tabular}{lcccccccc}
\hline\hline
Name  &  B  &  V  &  $\log_{10}{M_{*}/\msun}$ \\ \hline 
J0037$-$0942  &  -18.56  &  -18.72  &  $8.61^{+0.12}_{-0.17}$  \\
J0044$+$0113  &  -18.02  &  -18.45  &  $9.11^{+0.30}_{-0.16}$  \\
J0157$-$0056  &  -18.28  &  -19.07  &  $9.82^{+0.15}_{-0.17}$  \\
J0216$-$0813  &  -18.30  &  -18.68  &  $9.13^{+0.17}_{-0.21}$  \\
J0330$-$0020  &  -21.00  &  -21.20  &  $9.79^{+0.18}_{-0.19}$  \\
J0737$+$3216  &  -17.84  &  -18.19  &  $8.83^{+0.20}_{-0.18}$  \\
J0912$+$0029  &  -15.46  &  -16.09  &  $8.65^{+0.15}_{-0.16}$  \\
J0935$-$0003  &  -17.93  &  -18.26  &  $8.93^{+0.25}_{-0.16}$  \\
J0955$+$0101  &  -18.15  &  -18.70  &  $9.49^{+0.20}_{-0.15}$  \\
J0959$+$0410  &  -18.81  &  -19.46  &  $10.03^{+0.18}_{-0.14}$  \\
J1100$+$5329  &  -17.51  &  -17.63  &  $8.14^{+0.12}_{-0.12}$  \\
J1103$+$5322  &  -18.45  &  -19.20  &  $9.84^{+0.15}_{-0.17}$  \\
J1106$+$5228  &  -16.36  &  -16.86  &  $8.72^{+0.23}_{-0.12}$  \\
J1112$+$0826  &  -19.10  &  -19.24  &  $8.78^{+0.11}_{-0.17}$  \\
J1142$+$1001  &  -17.28  &  -17.44  &  $8.25^{+0.18}_{-0.17}$  \\
J1143$-$0144  &  -17.81  &  -18.18  &  $9.00^{+0.23}_{-0.15}$  \\
J1204$+$0358  &  -18.57  &  -18.76  &  $8.83^{+0.16}_{-0.21}$  \\
J1205$+$4910  &  -17.16  &  -17.30  &  $8.04^{+0.12}_{-0.16}$  \\
J1213$+$6708  &  -16.94  &  -17.17  &  $8.25^{+0.16}_{-0.23}$  \\
J1218$+$0830  &  -18.23  &  -18.68  &  $9.34^{+0.19}_{-0.14}$  \\
J1250$-$0135  &  -17.27  &  -17.96  &  $9.37^{+0.14}_{-0.15}$  \\
J1250$+$0523  &  -16.62  &  -17.51  &  $9.46^{+0.15}_{-0.15}$  \\
J1306$+$0600  &  -17.39  &  -17.61  &  $8.31^{+0.16}_{-0.18}$  \\
J1313$+$4615  &  -18.67  &  -19.12  &  $9.47^{+0.18}_{-0.18}$  \\
J1318$-$0313  &  -20.24  &  -20.45  &  $9.51^{+0.13}_{-0.27}$  \\
J1319$+$1504  &  -19.58  &  -20.06  &  $9.97^{+0.16}_{-0.18}$  \\
J1402$+$6321  &  -14.61  &  -14.73  &  $6.99^{+0.12}_{-0.13}$  \\
J1416$+$5136  &  -19.12  &  -19.39  &  $9.29^{+0.22}_{-0.19}$  \\
J1420$+$6019  &  -18.19  &  -18.75  &  $9.58^{+0.18}_{-0.13}$  \\
J1432$+$6317  &  -18.85  &  -19.07  &  $8.89^{+0.17}_{-0.19}$  \\
J1436$-$0000  &  -18.97  &  -19.27  &  $9.29^{+0.22}_{-0.17}$  \\
J1443$+$0304  &  -16.09  &  -16.24  &  $7.65^{+0.11}_{-0.25}$  \\
J1451$-$0239  &  -16.89  &  -17.08  &  $8.13^{+0.20}_{-0.15}$  \\
J1525$+$3327  &  -17.35  &  -17.74  &  $8.89^{+0.21}_{-0.14}$  \\
J1531$-$0105  &  -17.81  &  -17.97  &  $8.28^{+0.11}_{-0.19}$  \\
J1538$+$5817  &  -14.86  &  -14.97  &  $7.03^{+0.11}_{-0.12}$  \\
J1621$+$3931  &  -17.27  &  -17.50  &  $8.46^{+0.18}_{-0.22}$  \\
J1630$+$4520  &  -17.36  &  -17.69  &  $8.77^{+0.21}_{-0.17}$  \\
J1644$+$2625  &  -17.74  &  -18.06  &  $8.83^{+0.24}_{-0.16}$  \\
J1719$+$2939  &  -14.37  &  -14.47  &  $6.86^{+0.11}_{-0.12}$  \\
J2238$-$0754  &  -18.27  &  -18.42  &  $8.53^{+0.13}_{-0.21}$  \\
J2300$+$0022  &  -15.63  &  -15.89  &  $7.67^{+0.13}_{-0.23}$  \\
J2302$-$0840  &  -16.31  &  -16.69  &  $8.28^{+0.21}_{-0.28}$  \\
J2303$+$1422  &  -18.03  &  -18.33  &  $8.88^{+0.18}_{-0.19}$  \\
J2341$+$0000  &  -19.63  &  -20.08  &  $9.84^{+0.15}_{-0.17}$  \\
J2347$-$0005  &  -19.13  &  -19.62  &  $9.89^{+0.16}_{-0.15}$  \\
\hline
\end{tabular}
\end{center}
{\footnotesize Notes:
Absolute magnitudes are given in the Vega system. 
Stellar masses were inferred assuming a Kroupa IMF. 
The typical uncertainty on the absolute magnitudes is 0.17~mag, 
(0.1 statistical, 0.14 systematic) with an 
additional 0.26~mag systematic error (which does not affect the color) 
due to the uncertain lens mass density profile slope. 
This is also the source of a systematic error of 0.1~dex 
on the stellar mass, which is included in the given error bars.}
\end{table}

%% file: tab3.tex
\begin{table*}
\begin{center}
\caption{Model transformations and hard limits\label{tab:models}}
\begin{tabular}{lllll}
\hline\hline
       & Parameter                               & Prior                & Transformation        &  Limits         \\ \hline
\multirow{4}{*}{Lens mass profile}
       & Axis ratio $q=b/a$                      & Uniform              & Linear                &  $[0,1]$        \\
       & Velocity dispersion $\sigmaSIE$ (km s$^{-1}$)  & Uniform              & Linear                &  $[150,400]$    \\
       & Orientation angle $\thetad$ (rad)       & Uniform              & Linear                &  $[0,\pi]$      \\
       & Position $\xd,\yd$ (arcsec)             & Gaussian             & Error function        &  $[-1,1]$       \\
\hline
\multirow{5}{*}{Source surface brightness profile} 
       & Position angle (rad)                    & Uniform              & Linear                & $[0,\pi]$       \\
       & $cos(i)=q$                              & Uniform              & Linear                & $[0,1]$         \\
       & Effective radius $\Reff$ (arcsec)       & Uniform              & Linear                & $[10^{-3},100]$ \\
       & Magnitude $m$                           & Power law, index $k$ & Power law , index $k$ & $[10,30]$       \\
       & Sersic index $n$                        & Uniform              & Linear                & $[0.2, 10]$     \\
\hline
\end{tabular}
\end{center}
\end{table*}

%% file: tab4.tex
\begin{table}
\begin{center}
\caption{Mean parameter difference (inferred -- true)
and associated rms scatter for simulated non-lensed galaxies\label{tab:nonsim}}
\begin{tabular}{lll}
\hline\hline
Parameter                       & \klens models                 &     \galfit models              \\ \hline
$\langle \Delta m\rangle$       & $(118\pm1)\times10^{-4}$      &     $(-76.0\pm0.6)\times10^{-4}$  \\
$\sigma_m$                      & 0.027                         &     0.016                       \\
$\langle \Delta \Reff \rangle$  & $(-730''\pm7)\times10^{-5}$   &     $(546''\pm4)\times10^{-5}$  \\
$\sigma_r$                      & 0.019                         &     0.012                       \\
$\langle \Delta n\rangle$       & $(-350\pm3)\times10^{-4}$     &     $(159\pm2)\times10^{-4}$    \\
$\sigma_n$                      & 0.079                         &     0.041                       \\
\hline
\end{tabular}
\end{center}
\end{table}

%% file: tab5.tex
\begin{table}
\begin{center}
\caption{\klens\ parameter recovery in
a simulated lens sample -- common lens and source parameters \label{tab:synfix}}
\begin{tabular}{llll}
\hline\hline
Parameter          & True value  & Inferred value      & r.m.s. scatter \\ \hline
$q$                & 0.77        & $0.76044\pm0.00006$ & 0.063       \\
$\sigmaSIE$ (km s$^{-1}$) & 270         & $269.961\pm0.007$   & 7.294       \\
$\thetad$ (rad)    & 1.75        & $1.7368\pm0.0002$   & 0.221       \\
$\thetas$ (rad)    & 1.75        & $1.7288\pm0.0003$   & 0.293       \\
$\cos(i)$          & 0.6         & $0.58685\pm0.00004$ & 0.044       \\
\hline
\end{tabular}
\end{center}
\end{table}       

%% file: tab6.tex
\begin{table}
\begin{center}
\caption{\klens\ parameter recovery in 
a simulated lens sample -- variable source parameters \label{tab:synthetic}}
\begin{tabular}{lccc}
\hline\hline
               Parameter   & True value & Inferred value      & r.m.s. scatter \\ \hline
\multirow{3}{*}{effective radius} 
                           & $0.05''$   & $0.0543''\pm0.0006$ & $0.011''$   \\
                           & $0.1''$    & $0.1023''\pm0.0009$ & $0.016''$   \\
                           & $0.2''$    & $0.195'' \pm0.001$  & $0.022''$   \\
\hline
\multirow{3}{*}{magnitude} 
                           & $23$       & $23.01\pm0.02$      & $0.107$     \\
                           & $24$       & $24.00\pm0.01$      & $0.062$     \\
                           & $25$       & $24.95\pm0.02$      & $0.265$     \\
\hline
\multirow{3}{*}{Sersic index} 
                           & $0.5$      & $0.62\pm0.02$       & $0.354$     \\
                           & $1.0$      & $1.05\pm0.01$       & $0.247$     \\
                           & $2.0$      & $1.98\pm0.02$       & $0.291$     \\
\hline
\end{tabular}
\end{center}
\end{table}